\documentclass{JHEP3}
\usepackage{latexsym}
\usepackage{graphics}
\usepackage{epsfig}
\usepackage{multirow}
\usepackage{amssymb,amsmath}

\newcommand{\ie}{\emph{i.e.}~}

\title{A Consistent Prescription for the Production Involving  Massive Quarks in Hadron Collisions}

\author{Borut Paul Kersevan$^{a,b}$ \\ $^a$ Jozef Stefan Institute, Jamova 39, SI-1000 Ljubljana, Slovenia \\
$^b$  Faculty of Mathematics and Physics,University of Ljubljana,\\ Jadranska 19a, SI-1000 Ljubljana, Slovenia\\
E-mail: \email{borut.kersevan@ijs.si}}
\author{Ian Hinchliffe\thanks{Work supported  by the Director, Office of Science, Office of
High Energy Physics, of the U.S.\ Department of Energy under Contract
DE-AC02-05CH11231.}\\ Lawrence Berkeley National Laboratory, Berkeley, CA , 94720 USA.\\
E-mail: \email{I\_Hinchliffe@lbl.gov} }

\abstract{
This paper addresses the issue of production of charm or bottom quarks
in association with a high $p_T$ process in hadron hadron
collision. These quarks can be produced either as part of the hard
scattering process or as a remnant from the structure functions. The
latter sums terms of the type $(\alpha_s log(p_T/m_q))^n$. If
structure functions of charm or bottom quarks are used together with a
hard process which also allows production of these quarks double
counting occurs. This paper describes the correct procedure and
provides two examples of its implimentation in single top and
Drell-Yan at the LHC.
}


\preprint{hep-ph/06xxxx}

\begin{document}

\section{Introduction}

The production of states by a hard scattering QCD process in hadron collision is
described by the parton model which separates it into perturbative (calculable)
process and a soft non-perturbative piece. Consider the production of a top
quark pair via the partonic process $gg\to t\overline{t}$. There are many gluons
present in the process but the only top quarks arise from the hard scatter
itself. Top quarks can also be produced singly via the process $gb\to W^- t$. In
order for this to occur the incoming bottom quark is viewed as a constituent
(parton) of the incoming hadron. Alternatively one could begin with a two gluon
initial state and consider the hard process as $gg\to t\overline{b}W^-$. These
two processes cannot be added as the QCD approximation that produces the $b$
parton in the first case is partially accounted for by the latter process. A
careful separation of ``hard'' and ``soft'' components is needed so that a
consistent result can be obtained. The rest of this paper demonstrates such a
separation. The rest of this section is concerned with introducing the
formalism.  Section 2 shows explicitly how to relate processes with one more (or
less) parton in the hard scattering. Section 3 presents some explicit examples
of the Monte-Carlo implementation of this formalism.  Finally some conclusions
are drawn.

The QCD-based parton model is based on the factorization theorems 
\cite{Altarelli:1977zs,Collins:1977iv,Collins:1981uw,Collins:1998rz,Olness:1989ke}
according to which the squared amplitude for a process $\rm A,B \to X$ can be
decomposed into a ``hard'' and ``soft'' (or alternatively denoted ``short'' and
``long distance'') parts:
\begin{eqnarray}
|\mathcal{M}_{AB \to X}|^2 & = & \sum_{a,b} f_{a/A} \otimes H_{ab \to X} \otimes
f_{b/B} \nonumber \\  & = & \sum_{a,b} \int \frac{d\xi_a}{\xi_a} \int \frac{d\xi_b}{\xi_b} 
f_{a/A}(\xi_a,\mu_F)\, f_{b/B}(\xi_b,\mu_F)\, H_{ab \to X}(\xi_a,\xi_b,\mu_F\ldots),
\label{e:fact}
\end{eqnarray}
with $\rm a,b$  labeling the incoming partons which have to be summed
over and 
$\rm H(ab \to X)$ denoting the hard ('short time') part of the squared 
amplitude. The soft contributions are absorbed into the parton distribution
functions $\rm f_{i/I}(\xi_i,\mu_F)$ with $\rm \mu_F$ being the (factorization)
scale at which the two parts were separated\footnote{In case all partons are
considered massless the flux factor in the partonic cross-section expression is 
$\rm \hat{s}=\xi_a \xi_b (2s)$ with $(2s)$ being the hadronic flux 
and the Eq. \ref{e:fact} results in the  common expression $\rm \sigma_{AB \to X} 
= \int {d\xi_a}  \int {d\xi_b} f_{a/A}(\xi_a,\mu_F)\, f_{b/B}(\xi_b,\mu_F)\, 
\sigma^{\rm hard}_{ab \to X} (\hat{s},\mu_F)$}. 
More explicitly,  the
above theorem states that the collinear (mass) singularities have to be
isolated/subtracted from the hard process amplitudes and reabsorbed into the parton
distribution functions 
\cite{Collins:1981uw,Collins:1998rz,Olness:1989ke,Aivazis:1993kh,Aivazis:1993pi};
all other singularities (UV, soft IR) appearing in the perturbative calculation of the
 hard process either cancel or 
are handled by renormalization
procedures. It has to be stressed at this point that the renormalization/regularization scheme
used in subtracting the UV singularities in turn dictates the precise form of
the evolution (DGLAP) equations of the parton distribution functions
 \cite{Altarelli:1977zs,Collins:1981uw}
\begin{equation}
\frac{d}{d\ln \mu_F^2} f_{i/I}(z,\mu_F) = \frac{\alpha_s(\mu_F)}{2\pi} \sum_j
\int\limits_z^1 \frac{d\xi}{\xi} 
P_{j \to i}(\frac{z}{\xi},\alpha_s(\mu_F))\, f_{j/I}(\xi,\mu_F) ,
\label{e:dglap}
\end{equation}
where $\rm P_{j \to i}$ denote the usual DGLAP evolution kernels and $\rm I$
describes either a parton or a hadron. 

An elegant way of isolating the mass singularities in perturbative calculations
is found by observing that the pQCD squared amplitude $\rm |\mathcal{M}_{ab \to
X}|^2$ involving initial state partons $\rm a,b$ is subject to the same
factorization theorem:
\begin{equation}
|\mathcal{M}_{ab \to X}|^2 = \sum_{c,d} f_{c/a} \otimes H_{cd \to X} \otimes
f_{d/b},
\label{e:par_fact}
\end{equation}
with the $\rm f_{i/j}$ representing the distribution function of the parton
$\rm i$ inside the parton $\rm j$. The above equation holds to any order in
perturbation theory. Consequently, since the $|\mathcal{M}_{ab \to X}|^2$ can be calculated to
any order by using the Feynman rules and the prescriptions for calculating the
$\rm f_{i/j}$ to high orders in $\alpha_s$ are also well established
\cite{Altarelli:1977zs,Kosower:2003np,Moch:2002sn} one can use the procedure of 
\cite{Olness:1989ke,Aivazis:1993kh,Aivazis:1993pi} to
extract the $H_{cd \to X}$ to the chosen order.

At zero-th order in $\alpha_s$:
\begin{equation}
f_{i/j}^{(0)}(\xi) = \delta_j^i \delta(\xi - 1)
\label{e:fij0}
\end{equation}
and hence:
\begin{equation}
|\mathcal{M}_{ab \to X}^{(0)}|^2 = H_{ab \to X}^{(0)}.
\label{e:hij0}
\end{equation}
Subsequently, at first order in $\alpha_s$:
\begin{equation}
f_{i/j}(\xi) = f_{i/j}^{(0)}(\xi) + f_{i/j}^{(1)}(\xi),
\label{e:fij1} 
\end{equation}
and thus at this order:
\begin{equation}
|\mathcal{M}_{ab \to X}^{(1)}|^2 = H_{ab \to X}^{(1)} + \sum_{c} f_{c/a}^{(1)} \otimes H_{cb
\to X}^{(0)} + \sum_{d} H_{ad \to X}^{(0)} \otimes f_{d/b}^{(1)},
\label{e:hij1}
\end{equation}
where formally the virtuality $\rm \mu^2$ of the particles $c$ or $d$ (which is
in turn proportional to the $(p_T)^2$ of the particles $\bar{c}$ or $\bar{d}$,
see Fig. \ref{f:pdf}) is used as the factorization measure with the limit $\rm
\mu_F^2$: The hard part $\rm H_{ab \to X}^{(1)}$ includes the cases $\rm \mu^2
\geq \mu_F^2$ and the soft one the cases $\rm \mu^2 \leq \mu_F^2$. Note that in
the above equation, the phase space integration over $\bar{c}/\bar{d}$ particles has
already been performed (resulting in the convolution integral) and thus the
phase space for the final state particles for the
soft part formally involves one less particle.
\FIGURE{
     \epsfig{file=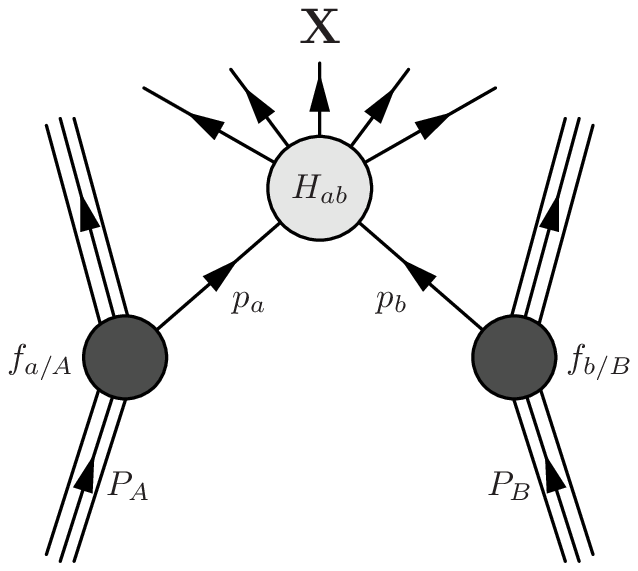,width=4.6cm}\parbox{0.3cm}{\vskip -3.5cm \Large$+$}
     \epsfig{file=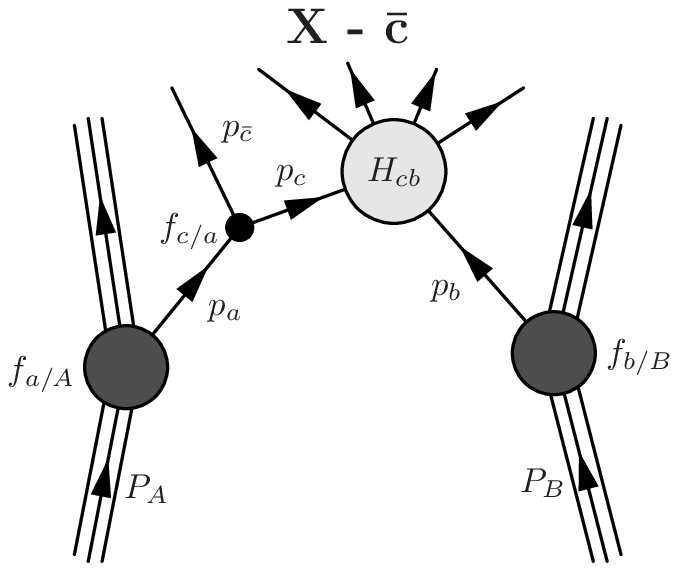,width=4.6cm}\parbox{0.3cm}{\vskip -3.5cm \Large$+$}
     \epsfig{file=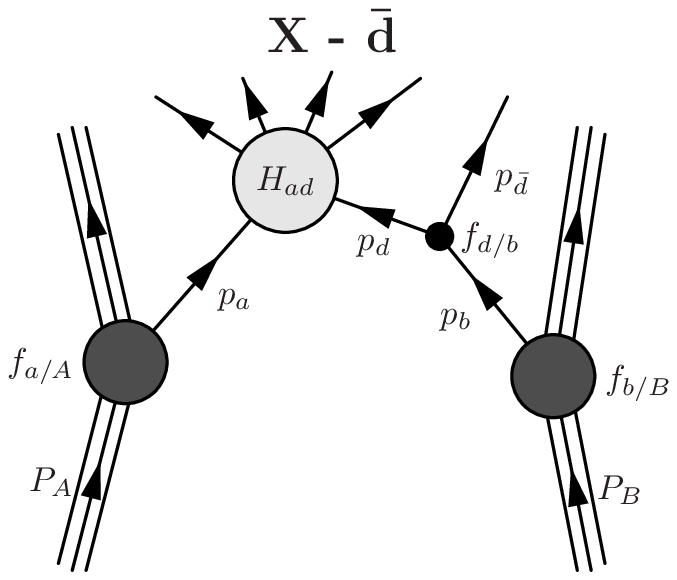,width=4.6cm}
\caption{
The diagrammatic representation of the method applied in isolating the
soft (collinear) terms in propagators corresponding to virtual particles $\rm
p_c$ and/or $\rm p_d$ when calculating the $|\mathcal{M}_{ab \to
X}^{(1)}|^2$. Note that in factorization theorem the particles incoming into the
hard part of the probability amplitude are considered to be on-shell.
\label{f:pdf}} 
}

The above equation can be inverted to give:
\begin{equation}
H_{ab \to X}^{(1)} = |\mathcal{M}_{ab \to X}^{(1)}|^2 - \sum_{c} f_{c/a}^{(1)} \otimes
|\mathcal{M}_{cb \to X}^{(0)}|^2 - \sum_{d} |\mathcal{M}_{ad \to X}^{(0)}|^2 \otimes f_{d/b}^{(1)}
\label{e:subt}
\end{equation}
This can be extended to  higher  orders in perturbation theory. It should be
emphasized that the presence of the subtraction terms in the above
Eq. \ref{e:subt} prevents double counting when
performing the perturbative cross-section calculation since the collinear
effects present in the $\rm |\mathcal{M}_{ab \to X}^{(1)}|^2$ are removed and
re-summed in the parton distribution functions $\rm f_{a/A}$ of the initial
hadrons.

After the perturbative expansion of $H_{ab \to X} = H_{ab \to X}^{(0)}
+ H_{ab \to X}^{(1)}$, given by Eq. \ref{e:hij0} and Eq. \ref{e:subt}, is inserted
into the cross-section expression (Eq. \ref{e:fact}) one thus obtains the formula:
\begin{equation}
|\mathcal{M}_{AB \to X}|^2 = |\mathcal{M}_{AB \to X}^{(0)}|^2 + 
|\mathcal{M}_{AB \to X}^{(1)}|^2 - |\mathcal{M}_{AB \to X}|^2_{\mathrm{s}},
\label{e:pertex}
\end{equation} 
with the subtraction terms given by:
\begin{equation}
|\mathcal{M}_{AB \to X}|^2_{\mathrm{s}}= 
\sum_{a,b} f_{a/A} \otimes \sum_{c} f_{c/a}^{(1)} \otimes H_{cb\to X}^{(0)} 
\otimes f_{b/B} + \sum_{a,b} f_{a/A} \otimes \sum_{d} H_{ad \to X}^{(0)} \otimes f_{d/b}^{(1)} \otimes f_{b/B}.
\label{e:showm}
\end{equation}

For further discussion on the 'double counting' issues it is illuminating to calculate the 
$f_{i/j}$ at scale $\mu_F$ up to the order of $\alpha_s$. 
One starts by writing down 
the perturbative expansion of the  evolution kernels in the DGLAP equations:
\begin{equation}
P_{j \to i}(\xi,\alpha_s(\mu_F))\, = P_{j \to i}^{(0)}(\xi) +
\left(\frac{\alpha_s(\mu_F)}{2\pi}\right) P_{j \to i}^{(1)}(\xi) + \ldots
\end{equation}
and observing that the evolution increases the order of $\rm f_{i/j}^{(n)}$ by
one. This can  explicitly be seen by inserting the zero-th order $\rm f_{i/j}^{(0)}$
of Eq. \ref{e:fij0} into the Eq. \ref{e:dglap}, and increasing the factorization scale
from the lowest kinematic limit (mass $\rm m$ of the particle $\rm i$) to the
scale $\mu_F$, \ie  integrating over the range $\rm [m^2,\mu_F^2]$ 
and keeping only the terms up to the order of $\alpha_s$:
\begin{equation}
f_{i/j}(\xi,\mu_F) = f_{i/j}^{(0)}(\xi)+\frac{\alpha_s(\mu_F)}{2\pi} 
P_{j \to i}^{(0)}(\xi) \ln\left(\frac{\mu_F^2}{m^2}\right).
\label{e:collog}
\end{equation}
The above expression matches the perturbative expansion given by Eq. \ref{e:fij1} with the 
second term identified as the $\rm f_{i/j}^{(1)}$ parton distribution
function.

While the expression of Eq. \ref{e:pertex} can subsequently be used for
estimating the total cross-section of a given process one should take further
steps when dealing with the estimation of the differential cross-sections or
(equivalently) Monte-Carlo simulation. In a Monte-Carlo simulation the
DGLAP parton evolution, re-summed in the parton density functions $f_{i/I}$
(c.f. Eq. \ref{e:dglap}), is made explicit by evolving the factorization scale
from its initial value $\rm \mu_0$ to the lowest kinematic limit (or an imposed
cutoff). Identifying the evolving factorization scale with the virtuality $\rm
\mu^2$ of the incoming particle $\rm c$ the probability that the particle $\rm c$ will be
un-resolved into a particle $\rm a$ (or equivalently, that a particle $\rm a$
will branch and produce the particle $\rm c$ and an additional spectator
particle) is given by the Sudakov term (see e.g. \cite{Sjostrand:2001yu}):
\begin{equation}
S_a = \exp{\left\{- \int\limits_{\mu^2}^{\mu_0^2} \frac{d\mu'^2}{\mu'^2} 
\frac{\alpha_s(\mu'^2)}{2\pi} \times \sum\limits_{c} 
\int\limits_{\xi_c}^{1} \frac{dz}{z}  P_{a \to c}(z) 
\frac{f_{a/I}(\frac{\xi_c}{z},\mu'^2)}{f_{c/I}(\xi_c,\mu'^2)}
\right\}}.
\label{e:sudakov_gen}
\end{equation}
At each evolution step (branching) the number of particles is increased by one and
its contribution to the differential cross-section in terms of $\alpha_s$ is also increased by one. The described procedure is commonly known as (initial state) \emph{parton showering}. 
 
The (next-to-leading order) subtraction terms of Equation \ref{e:showm} thus compensate for the first branching 
in the backward evolution of the incoming partons participating in the (leading order) term 
$H_{ab \to X}^{(0)}$.
Note that in order to match the subtraction terms in Eq. \ref{e:subt} with the
first--order matrix element, the fraction $\rm \xi_c$ (or equivalently $\rm
\xi_d$) of the evolved parton is kept constant and the virtuality is 
\emph{decreased} corresponding to a 'backward' evolution in time from the
starting point of virtuality $\rm \mu^2_0$ of $\rm H_{ab \to X}^{(0)}$. to the
virtuality $\rm \mu^2$ of $\rm H_{ab \to X}^{(1)}$ (c.f. Figure \ref{f:pdf}). 
Writing the expressions of Equation \ref{e:showm} in differential form in $\rm \mu^2$ one thus gets for the first term:
\begin{eqnarray}
|\mathcal{M}_{AB \to X}|^2_{\mathrm{s}}=  \frac{d\mu^2}{\mu^2} 
 \sum_{a,b,c} \int \frac{d\xi_a}{\xi_a} \int  \frac{d\xi_b}{\xi_b} 
\int \frac{d\xi_c}{\xi_c} && \Biggl\{ f_{a/A}(\xi_a,\mu_0^2)\, 
\frac{\alpha_s(\mu^2_0)}{2\pi}  P_{a \to c}^{(0)}\left(\frac{\xi_c}{\xi_a}\right)\, 
 \nonumber \\  && H_{cb\to X}^{(0)}(\xi_c,\xi_b) \, f_{b/B}(\xi_b,\mu_0^2) + \ldots \Biggr\} 
\end{eqnarray}
and an equivalent expression can be obtained for the second term of Eq. \ref{e:showm}.
Using again the Eq. \ref{e:hij0}, multiplying by the flux factor 
${1}/{2 \sqrt{\lambda(s,m_A^2,m_B^2)}}$, given by the Lorentz
invariant function:
\begin{equation}
\lambda(s,m_1^2,m_2^2)=(s-(m_1+m_2)^2)(s-(m_1-m_2)^2)
\label{e:lambda}
\end{equation}
and integrating over the final state n-particle phase space denoted by 
$\rm \int d\Phi_X$, one obtains the first subtraction term:
\begin{eqnarray}
&& \frac{d\sigma^{(0)}_{s1}(AB \to X)}{d\xi_a d\xi_b d\mu^2 d\xi_c d\phi} =
\label{e:dshow1} \nonumber \\ 
&&\sum\limits_{a,b,c} 
\frac{\theta(\mu_0^2 - \mu^2)}{2 \sqrt{\lambda(s,m_A^2,m_B^2)}\xi_c \xi_b} 
\frac{\alpha_s(\mu^2_0)}{4\pi^2\, \mu^2} \frac{1}{\xi_a} f_{a/A}(\xi_a,\mu_0^2)\,
 P_{a \to c}^{(0)}\left(\frac{\xi_c}{\xi_a}\right)\,  f_{b/B}(\xi_b,\mu_0^2) \times \nonumber \\ 
&& \qquad \times \int |\mathcal{M}_{cb \to X}^{(0)}|^2(\xi_c,\xi_b) d\Phi_{X-\bar{c}} 
\end{eqnarray}
and equivalently also the second subtraction term by appropriate
replacements $a \to b$ and $c \to d$. The two derived equations correspond
to the expressions obtained by Chen,Collins \emph{et al.}
\cite{Collins:2000qd,Chen:2001nf,Chen:2001ci,Collins:2002ey}, derived
by the Sudakov exponent expansion.

In addition to writing the cross-sections in differential form, the integration
over an angle $\rm \phi$ was introduced, where the angle $\phi$ represents the
azimuthal angle of the spectator particle $\rm \bar{c}$ or $\rm \bar{d}$ and is in
effect a dummy quantity which nevertheless has to be sampled in a Monte--Carlo
simulation procedure. The notation $\rm d\Phi_{X-\bar{c}}$ denotes that the
final space integral does not contain the spectator particles $\rm \bar{c}$ or $\rm
\bar{d}$ since they are already accounted for in the $\rm (d\xi/\xi) d\mu^2 d\phi$
differential (see e.g. \cite{Altarelli:1977zs}). In contrast the first order matrix
element of Equation \ref{e:subt} is integrated over the full phase space X:
\begin{eqnarray} 
&& \frac{d\sigma^{(1)}(AB \to X)}{dx_a dx_b} = \label{e:xsec1}
 \nonumber \\ 
&& \sum\limits_{a,b}
\frac{1}{2 \sqrt{\lambda(s,m_A^2,m_B^2)}x_a x_b} 
 f_{a/A}(x_a,\mu_0^2)\, f_{b/B}(x_b,\mu_0^2) \int 
|\mathcal{M}_{ab \to X}^{(1)}|^2(x_a,x_b)d\Phi_X ~,  
\end{eqnarray}
whereby the outstanding issue is the kinematic translation between $\rm n-1$ 
($\rm X-\bar{c}$) and $\rm n$ ($\rm X$) particle kinematics. Furthermore, one cannot
simply equate the variables $\rm \xi_{a,b}$ and $\rm x_{a,b}$, since e.g. in Eq.
\ref{e:dshow1} the $\rm \xi_{c,b}$ terms imply that the incoming particles $\rm c$ and
$\rm b$ are on shell in the matrix element calculation while in
Eq. \ref{e:xsec1} in contrast the particle $\rm a$ (as the 'parent' of particle 
$\rm c$) is the on-shell one. 

The mapping of the kinematic quantities between between the expressions of order
$\rm \alpha_S$ (\ie the pQCD derived expression of Eq. \ref{e:xsec1} and the
(showering) subtraction terms of Eq. \ref{e:dshow1}  ) needs a consistent 
and possibly a formally correct definition. In order to achieve this the 
 prescription developed by Collins \emph{et al.} of how to merge
non-perturbative (parton-shower) calculations with the leading order perturbative pQCD
calculations on the level of Monte-Carlo simulations
\cite{Collins:2000qd,Chen:2001nf,Chen:2001ci,Collins:2002ey} was implemented, 
which has been explicitly shown to reproduce the NLO $\rm \bar{MS}$ result for a set of processes.

Another outstanding issue is that in case of heavy quarks participating as the
initial state partons the (commonly used) approximation of treating the incoming
particles as massless can lead to a significant error. This fact, as well as the
the solution in terms of consistent treatment of the kinematics in terms of
light-cone variables, has been demonstrated in the ACOT prescriptons of how to
consistently introduce the factorization in case of non-negligible masses of the
colliding partons (e.g. heavy quarks)\cite{Aivazis:1993kh,Aivazis:1993pi}. The
ACOT prescription has in this paper been introduced in the formalism of
Monte-Carlo simulation by modifying the prescription of Collins \emph{et al.} 
accordingly.

In the Monte--Carlo generation steps one thus has to produce two classes of events; one 
class is derived from the leading order process with one branching produced by Sudakov parton
showering and the second class are events produced from the next-to-leading order hard process
calculation (\ie the pQCD calculation with the appropriate subtraction terms).
 
\section{Kinematic Issues}

\subsection{Phase-Space Transformation \label{s:npart}}
\FIGURE{
     \epsfig{file=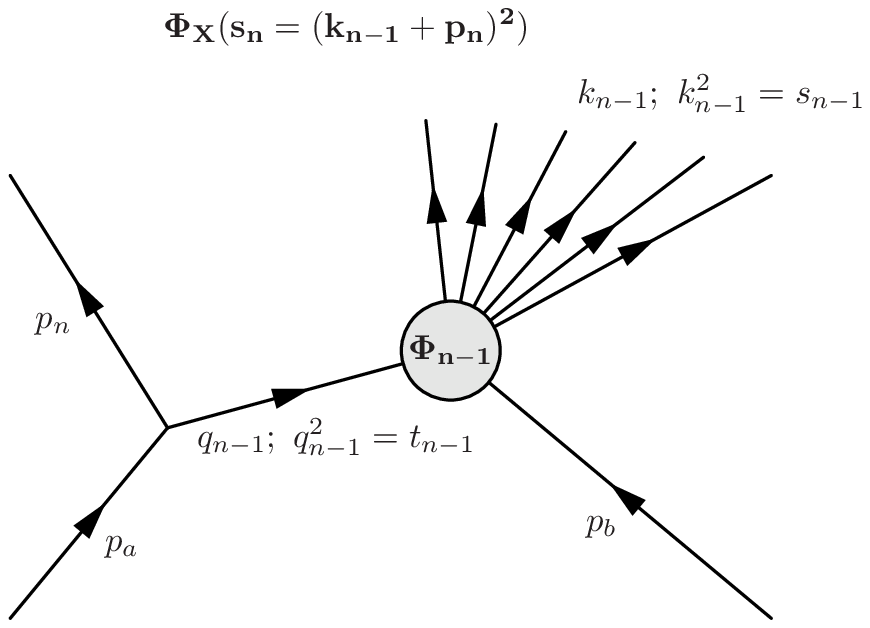,width=6.0cm}
\caption{The diagrammatic representation of the method applied in translating a
t-channel (space-like) split going from $n$ to $n-1$ particle phase space.
\label{f:kb}}}

The aim of this section is to derive generic expressions that transform the
kinematics from the 'hard' to the 'soft' (or showering) n-particle system,
\ie  split the 'hard' $\rm n$-particle phase space involving heavy quarks into
$\rm n-1$ 'hard' $ \oplus 1$ 'soft' particle phase space in order to perform
appropriate MC simulation (c.f. Figure \ref{f:kb}).  As already stated the
existing  prescriptions deal either with massive particles 
\cite{Aivazis:1993kh,Aivazis:1993pi} on the level of integrated cross-sections or
with explicit Monte-Carlo algorithms involving light ($\sim$ massless) particles (e.g.
\cite{Bengtsson:1987rw,Sjostrand:2001yu,Chen:2001ci,Catani:2001cc,Mrenna:2003if}
) while no generic combination of the two algorithms is available.

In order to accommodate the particle masses it is convenient to work in
light-cone coordinates $p^\mu=(p^+,\vec{p}^T,p^-)$ where $\rm p^\pm = \frac{1}{\sqrt{2}}(p^0 \pm p^3)$
and the remaining two coordinates are considered 'transverse' $\vec{p}^T$. 
The kinematic prescription of relating the hard $\rm n$-particle kinematics to 
the soft $\rm n-1$ case is as follows:
\begin{enumerate}
\item Incoming hadron A is moving in the $\rm z$ direction and hadron B in the $\rm -z$
direction, carrying momenta $\rm P_A$ and $\rm P_B$ with the center-of-mass
energy $\rm \sqrt{s}$, whereby one can neglect the hadron masses at LHC energies.
\item The incoming partons with momenta $\rm p_a$ and $\rm p_b$ have the
momentum fractions $\rm p_a^+ = x_a P_A^+$ and $\rm p_b^- = x_b P_B^-$
relative to the parent hadrons and the center-of-mass energy $\sqrt{s_n}$.
\item The split propagator (\ie particle $\rm c$) virtuality 
$p_c^2 = (p_a  - p_{\bar{c}})^2 = t_{n-1}$ (c.f. Fig. \ref{f:kb}) gives the $\rm
\mu^2$ value from $ t_{n-1} - m_{\bar{c}}^2 = -\mu^2 $. (c.f. Eq. \ref{e:dshow1}). 
\item The scale correspondence is given by 
$\rm s_{n-1} = \mu_0^2$ (c.f. Eq. \ref{e:dshow1}). 
\item All incoming and outgoing particles (partons) are on mass shell. 
\item The splitting parameter of the evolution kernel is 
$\rm z = \frac{\xi_c}{\xi_a}$ (already used in Eq. \ref{e:dshow1}). 
\item The rapidity $\rm y = \frac{1}{2} \ln\left( \frac{k_{n-1}^+}{k_{n-1}^-} \right)$ of
the subsystem  (c.f. Figure \ref{f:kb}) is preserved in the translation. 
\end{enumerate}
In order to further relate the $\rm n$ and $\rm n-1$ particle phase space for
hard and soft interpretation one can use the recursive t-channel splitting relation 
\cite{Kersevan:2004yh,KB}:
\begin{eqnarray}
&& \Phi_n(s_n,m_1,m_2,\ldots,m_n) =  \label{e:recurt} \nonumber \\
&& = \int\limits_{(\sum_{i=1}^{n-1}m_i)^2}^{(\sqrt{s_n} - m_n)^2} 
\frac{ds_{n-1} }{4\sqrt{\lambda(s_n,m_{a}^2,m_{b}^2)}} \int\limits_0^{2\pi} d\varphi_n^*
\int\limits_{t_{n-1}^{-}}^{t_{n-1}^{+}} dt_{n-1}\: 
 \Phi_{n-1}(s_{n-1},m_1,m_2,\ldots,m_{n-1}), 
\end{eqnarray}
where $\rm \varphi_n^* \equiv \phi$ of Eq. \ref{e:dshow1} and the limits
$t_{n-1}^{\pm}$ are given by analytic, albeit complex expressions
\cite{Kersevan:2004yh,KB}.  Using the above relation one can  introduce 
the $\rm X-\bar{c}$ particle phase space split into Equation \ref{e:xsec1} by
identifying $\rm p_{\bar{c}} \equiv p_n$ :
\begin{eqnarray} 
&& \frac{d\sigma^{(1)}(AB \to X)}{dx_a dx_b dt_{n-1} ds_{n-1} d\varphi^*} = 
\label{e:xsec1t} \nonumber \\ 
&& \sum\limits_{a,b} 
\frac{1}{2 \sqrt{\lambda(s,m_A^2,m_B^2)}x_a x_b} 
 f_{a/A}(x_a,s_{n-1})\, f_{b/B}(x_b,s_{n-1}) \times \nonumber \\ 
&& \qquad \times \int  |\mathcal{M}_{ab \to X}^{(1)}|^2(x_a,x_b)
\frac{1}{4\sqrt{\lambda(s_n,m_{a}^2,m_{b}^2)}} d\Phi_{X-\bar{c}} ~,  
\end{eqnarray}
Using the translation prescriptions introduced above,  the remaining
issue
is the  relation of  the variables $\rm \xi_a,\xi_c,\xi_b$ with the variables
$x_a,x_b,s_{n-1}$. The requirement ({\bf 2}) in the list above ensures 
$\rm  \xi_a \equiv x_a$  since both particles in question are incoming partons
originating in hadron A. The
remaining relations between  $\rm \xi_c,\xi_b$ and $x_b,s_{n-1}$ are then given
by energy and rapidity conservation requirements. The derived relations are explicitly listed in Appendix
\ref{app:conditions}.

Combining all the derived rules for kinematic
translation one finally obtains a transformation of Eq. \ref{e:dshow1}:
\begin{eqnarray}
&&\frac{d\sigma^{(0)}_{s1}(AB \to X)}{dx_a dx_b dt_{n-1} ds_{n-1} d\varphi^*} = \nonumber \\
&& = \mathcal{J}\frac{(\xi_c,\xi_b)}{(s_{n-1},x_b)} \left. 
\frac{d\sigma^{(0)}_{s1}(AB \to X)}{d\xi_a d\xi_b d\mu^2 d\xi_c d\phi} 
\right|_{\xi_a \to x_a,\, \mu^2 \to -(t_{n-1}-m_{\bar{c}}^2),\, \phi \to \varphi^*,\, (\xi_c,\xi_b)
\to (s_{n-1},x_b)},
\end{eqnarray}
where $\mathcal{J}\frac{(\bar{\tau},\bar{y})}{(s_{n-1},x_b)}$ is the Jacobian of the transformation derived
 in the Appendix \ref{app:conditions}.

An issue which deserves special consideration is the prescribed
substitution $ t_{n-1} - m_{\bar{c}}^2 = -\mu^2 $, where the split
particle virtuality $\mu^2$ is the propagator virtuality shifted by
the spectator mass $ m_{\bar{c}}^2$. This prescription differs from
the one of Collins \cite{Collins:2000qd}, where the relation is
directly $ t_{n-1} = -\mu^2$ and the spectator (propagator) mass shift
is omitted. The reason for this modification is clear when
one notes that the phase space limits $[t_{n-1}^{-},t_{n-1}^{+}]$
of the $t_{n-1}$ parameter are functions of $s_n$ and invariant masses
of the objects (particles) participating in a t-channel split of
Equation \ref{e:recurt} \cite{Kersevan:2004yh,KB} and thus do not
match the simple limits $[m_{\bar{c}}^2,s_{n-1}]$ of the virtuality
$\mu^2$ in the equation \ref{e:dshow1}. Indeed, studies have shown
that the presence of the cutoff $\theta(s_{n-1} + t_{n-1})$ does not
provide a sufficient solution since it only sets the upper integration
limit to $t_{n-1}^{+} \to -s_{n-1}$ while the lower limit
$t_{n-1}^{-}$ can in certain instances be even smaller than
$m_{\bar{c}}^2$. The reproduction of a logarithmic term of
the collinear singularity $\ln\left(\frac{\mu_F^2}{m^2}\right)$ (Eq. \ref{e:collog}) with 
$\mu_F^2 =\mu_0^2 = s_{n-1}$:
\begin{equation}
\int\limits_{t_{n-1}^{-}}^{t_{n-1}^{+}} \frac{\theta(s_{n-1} + t_{n-1})}{t_{n-1}} dt_{n-1}\: =
\ln\left(\frac{-s_{n-1}}{t_{n-1}^{-}}\right)
\neq  \ln\left(\frac{s_{n-1}}{m_{\bar{c}}^2}\right)
\end{equation}
is thus not satisfied in when using $ t_{n-1} = -\mu^2$. In
order to resolve this issue one needs to return  to the basics of the
factorization procedure, where the actual collinear singularity
(\ie \  the corresponding logarithmic term) is isolated from the
(integrated) hard process cross-section (for a nice example with
massive particles see {\it e.g.}  \cite{Aivazis:1993pi}) and these
logarithmic terms match with the required collinear logarithm
$\ln\left(\frac{\mu_F^2}{m^2}\right)$ only in the high $s_n$ (hard
center-of-mass) limit. The logarithmic collinear terms can
subsequently be traced back to the propagator integral:
\begin{equation}
\int\limits_{t_{n-1}^{-}}^{t_{n-1}^{+}} \frac{dt_{n-1}}{t_{n-1}-m_{\bar{c}}^2} =
\ln\left(\frac{t_{n-1}^{+}-m_{\bar{c}}^2}{t_{n-1}^{-}-m_{\bar{c}}^2}\right)
~~~\underrightarrow{s_n \to \infty} ~~~\ln\left(\frac{s_{n-1}}{m_{\bar{c}}^2}\right).
\label{e:shiftint} 
\end{equation}
The expression of Eq. \ref{e:shiftint} is indeed found to match the
logarithmic terms of \cite{Aivazis:1993pi}
exactly\footnote{Specifically the expressions of Eq. 17, page 14,
whereby one has to write down the explicit expressions for the limits
on $t_{n-1}$ and set the mass of the incoming particle corresponding
to the incoming gluon to zero in order to reproduce the kinematical
topology of the process studied in \cite{Aivazis:1993pi}.}. In order
to reproduce the collinear cutoff one thus has to put $-\mu^2 =
t_{n-1}-m_{\bar{c}}^2$ which, combined with the factorization cutoff
$\theta(\mu_F^2 - \mu^2) = \theta(s_{n-1} - \mu^2)$,  reproduces
the required logarithm in the high $s_n$ limit:
\begin{equation}
\int\limits_{-(t_{n-1}^{-}-m_{\bar{c}}^2)}^{-(t_{n-1}^{+}-m_{\bar{c}}^2)} \frac{\theta(s_{n-1} - \mu^2)}{\mu^2} d\mu^2\: =
\ln\left(\frac{s_{n-1}}{-(t_{n-1}^{-}-m_{\bar{c}}^2)}\right)
~~~\underrightarrow{s_n \to \infty} ~~~\ln\left(\frac{s_{n-1}}{m_{\bar{c}}^2}\right).
\end{equation}
\clearpage

\subsection{Monte-Carlo Generation Steps}

\FIGURE{
     \epsfig{file=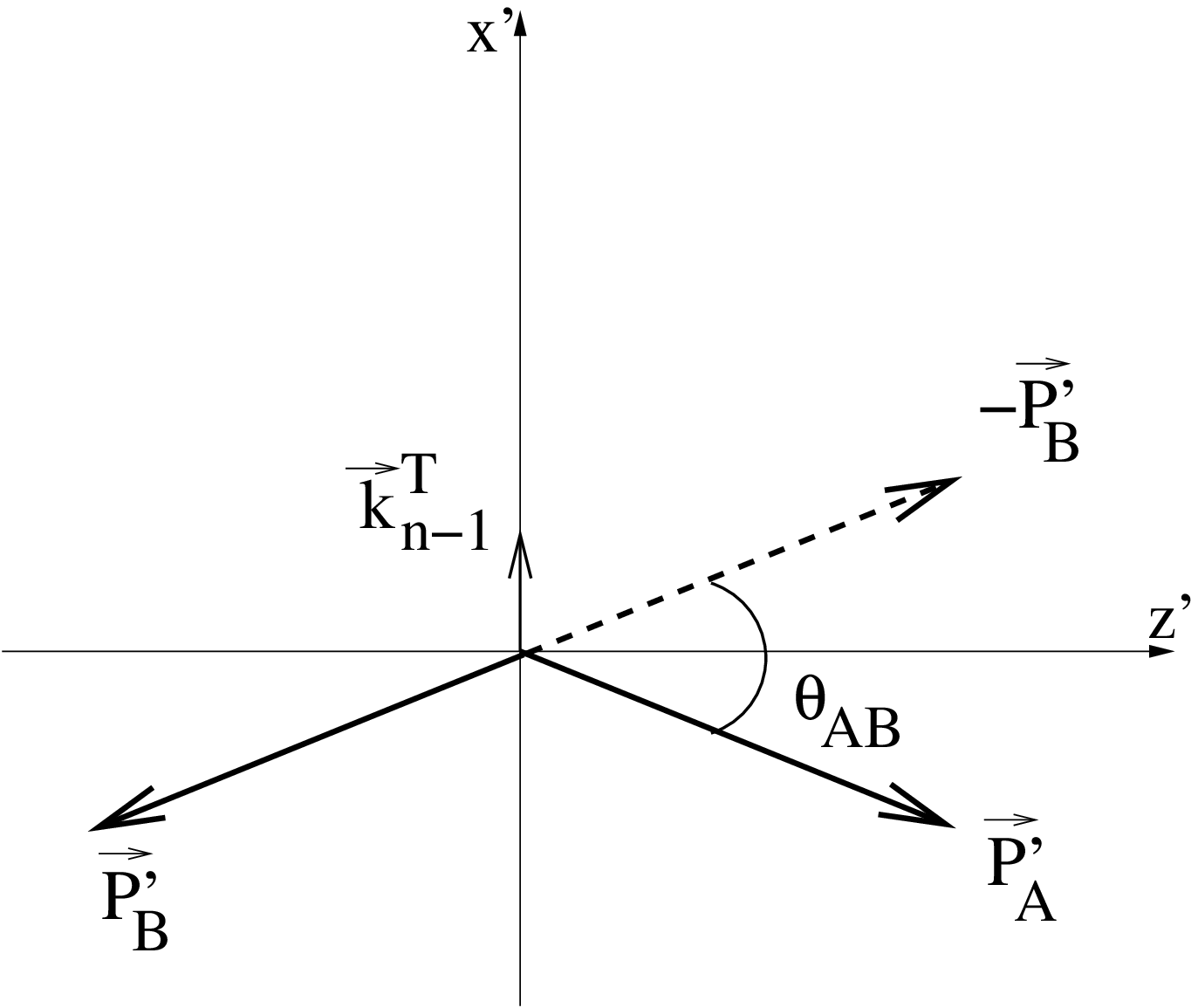,width=6.0cm}
\caption{ The definition of the reference Collins-Soper frame as the rest frame of
the subsystem $\rm s_{n-1}$ with the transverse component $\vec{k}^T_{n-1}$
oriented in the $x'-axis$ direction.
\label{f:colsop}} 
}

A full Monte--Carlo event generation  results using the following prescription:

\begin{enumerate}
\item \parbox[t]{7.2cm}{Generate the particle momenta and corresponding phase space weight for the
n-particle final state $\rm X$ and compute the full weight corresponding to the
process of Eq. \ref{e:xsec1}.}
\item \parbox[t]{7.2cm}{Re-calculate the kinematic quantities of the $\rm n \to (n-1) \oplus 1$
transformation as described above.}
\item \parbox[t]{7.2cm}{Boost the whole system into the Collins-Soper frame \cite{Collins:1977iv} of the
$\rm n-1$ subsystem,
 \ie the $s_{n-1}$ center of mass frame where the angle
between the boosted hadron momenta $P_A'$ or $-P_B'$ and $\rm z'$-axis now equals}\\ 
$\tan(\frac{1}{2} \theta_{AB}) = |\vec{k}^T_{n-1}|/{\sqrt{s_{n-1}}} =
|\vec{p}^T_{\bar{c}}|/{\sqrt{s_{n-1}}}$ (c.f. Fig. \ref{f:colsop}). In other
words, this transformation manifestly puts the transverse contribution due to
the induced virtuality into the hadron momenta directed perpendicularly to the  
$\rm z'$-axis. One can thus eliminate this virtuality by shifting the hadron
momenta to the $\rm z'$-axis while preserving the center-of mass energy
$s_{n-1}$.
\item The $n-1$ (hard) system corresponding to the Eq. \ref{e:dshow1} is then 
achieved by eliminating the particle $\bar{c}$ and boosting the remaining particles in the 
 $\rm z$-axis direction with the boost value of:
\begin{equation}
\beta = \frac{\xi_c ( \xi_c \xi_b s + m_b^2) - \xi_b ( \xi_c \xi_b s + m_c^2) }
{\xi_c ( \xi_c \xi_b s + m_b^2) + \xi_b ( \xi_c \xi_b s + m_c^2) }
\end{equation}

 in order to restore the sub-system rapidity $\rm y$. 
\item Since boosts do not change the phase space weight the necessary
modification consists of multiplying the phase space weight by:
\begin{equation}
4\sqrt{\lambda(s_n,m_{a}^2,m_{b}^2)}\theta(s_{n-1} + t_{n-1}) 
\mathcal{J}\frac{(\xi_c,\xi_b)}{(s_{n-1},x_b)}
\end{equation}
and then obtaining the first subtraction weight by putting the reconstructed
momenta into the Eq. \ref{e:dshow1}.
\item An analogous procedure can be repeated to obtain the alternate kinematic configuration
(with the parton evolution of the other incoming particle) and the second subtraction weight.
\item The final weight, after performing both subtractions is then passed to the
event unweighting procedure.
\end{enumerate}

As Chen, Collins \emph{et al.} pointed out, 
\cite{Collins:2000qd,Chen:2001nf,Chen:2001ci,Collins:2002ey}, the described
procedure of subtraction is not equivalent to the standard subtraction schemes
(e.g. $\rm \overline{MS}$) used to obtain the cross-sections for specific processes
(see e.g. \cite{Rijken:1994sh,Sutton:1991ay}) and hence the data-fitted parton distribution
functions with the corresponding evolution kernels (for example the widely used
CTEQ5 PDF-s \cite{Lai:1999wy}). The relation between the procedure applied
derived by Chen, Collins \emph{et al.} for the massless case and (with inclusion
of parton masses) applied above states that the correspondence between the 
$\rm \overline{MS}$ scheme and the applied $\rm JCC$ scheme is given by relatively
simple relations; an example for the expression of the quark $\rm i$ distribution
function involves a convolution of the gluon $g$ distribution function and
the $g \to i \bar{i}$ splitting kernel $P_{g \to i \bar{i}}(z)$:
\begin{eqnarray}
z\,f^{\mathrm{JCC}}_{i/I}(z,\mu^2) &=& z\,f^{\mathrm{\overline{MS}}}_{i/I}(z,\mu^2) \\
\notag &+& \frac{\alpha_s(\mu^2)}{2\pi} \int\limits_z^1 \,d\xi \frac{z}{\xi} 
  f^{\mathrm{\overline{MS}}}_{g/I}(\xi,\mu^2) 
\left[ P_{g \to i \bar{i}}(\frac{z}{\xi}) \ln\left(1 -  \frac{z}{\xi}\right) +
\frac{z}{\xi}\left(1 -  \frac{z}{\xi}\right) \right] \\ \notag
&+& \mathcal{O}\mathrm{(\text{first-order quark terms})}~+\mathcal{O}(\alpha_s^2)
\label{e:jcc}
\end{eqnarray}
These new distributions can in a reasonably straightforward manner be obtained by numerical 
integration. 

In order to complement the subtracted process one also has to generate a
parton-shower evolved zero-th order process with $\rm (n-1)$ particles 
participating in the hard process and an additional particle added by 'soft'
evolution of the incoming particles. Since one is interested in the heavy
initial state quarks this implies that one has to 'unresolve' one of
the initial quarks back to a gluon, whereby an additional (anti) quark is
added. The procedure to achieve this is straightforward
\cite{Collins:2000qd,Chen:2001nf,Chen:2001ci,Collins:2002ey} and complementary
to the procedure described above, \ie \ one has to perform the following steps in
the Monte-Carlo algorithm:
\begin{enumerate}
\item Generate the particles corresponding to the $\rm (n-1)$ phase space
topology, along with the momentum fractions $\xi_c$ and $\xi_b$ of the incoming
particles in the sense of light-cone components. Consequently, the invariant
mass of the hard system is $s_{n-1}$ and the rapidity $y$ is given by Equation
\ref{e:ynm1}.
\item Generate a virtuality $\rm \mu^2$ of the incoming heavy quark $c$, a
longitudinal splitting fraction $\rm z$ for the branching of gluon $a$ into the
$c \bar{c}$ pair and an azimuthal angle $\rm \phi$ of the branching system. All
the values are sampled from the Sudakov-type distribution:
\begin{equation}
S_a = \exp{\left\{- \int\limits_{\mu^2}^{\mu_0^2} \frac{d\mu'^2}{\mu'^2} 
\frac{\alpha_s(\mu'^2)}{2\pi} \times 
\int\limits_{\xi_c}^{1} \frac{dz}{z}  P_{a \to c}(z) 
\frac{f_{a/I}(\frac{\xi_c}{z},\mu'^2)}{f_{c/I}(\xi_c,\mu'^2)}
\right\}}.
\label{e:sudakov}
\end{equation}
In case there are two quarks in the initial state that can evolve back to gluon
and give the contribution of the same order (like e.g. $\rm b \bar{b} \to Z^0$
process) both virtualities are sampled and the quark with the higher one is
chosen to evolve. 
\end{enumerate}
Subsequently, the four-momenta of the participating particles are reconstructed
requiring that the subsystem invariant mass $s_{n-1}$ and the
rapidity $y$ are preserved; the construction is of course identical to the  
one used in the $\rm n \to (n-1) \oplus 1$ transformation given in Section \ref{s:npart}.
A point to stress is that a kinematic limitation arises on the allowed $\rm x_a$
and thus $\rm z = \frac{\xi_c}{x_a}$ values due to the requirement $|\vec{p}_{\bar{c}}^T|^2 \geq 0$. The
latter condition gives the minimal value of the invariant mass of the n-particle
system $\rm s_{n} = x_a x_b s + m_b^2 $ (taking into account that one of the incoming
particles is a gluon, $ m_a=0$) with:
\begin{equation}
\left( x_a x_b s \right)_{\mathrm{min}} = \frac{(m_c^2 +
\mu^2)}{2} \left( \frac{(s_{n-1}+\mu^2 - m_b^2)}{\mu^2} + \sqrt{\frac{(s_{n-1}+\mu^2 - m_b^2)^2}{\mu^4} + \frac{4 m_b^2}{\mu^2}} \right),
\end{equation}
which combines with the rapidity $\rm y$ conservation requirement into:
\begin{equation}
x_a^2 \geq \frac{\left( x_a x_b s \right)_{\mathrm{min}}^2 \left( \left( x_a x_b
s \right)_{\mathrm{min}} - (m_c^2 + \mu^2) \right)}{s \left[
(s_{n-1}+\mu^2)e^{-2y} \left( x_a x_b s \right)_{\mathrm{min}} -  m_b^2 (m_c^2 +
\mu^2) \right]}.
\end{equation}
In the massless limit the above expression translates back into the requirement
$x_a \geq \xi_c$ or equivalently $z \leq 1$. In practice (\ie Monte-Carlo
generation) this thus means that a certain fraction of generated topologies have
to be rejected and/or re-generated until the above conditions are met (or
equivalently, that the $z$ (or $\rm x_a$) sampling limits have to be shifted).
\clearpage

\section{Examples of the Algorithm Implementation}

Three examples of the procedure described in this paper have been developed: The
associated $Z^0 b$ production process and the 't-channel' and 'tW-channel'
single top production processes, both expected to be observed at the LHC. The
processes were implemented in the AcerMC Monte-Carlo generator
\cite{Kersevan:2002dd}. Due to the subtraction terms a fraction of event
candidates achieve negative sampling weights and unweighted events are produced
with weight values of $\pm 1$ using the standard unweighing procedures.

\subsection{Associated Drell-Yan and b-quark Production}

The (Drell-Yan) lepton pair production associated with one or more heavy quarks
represents an important irreducible background component in the Higgs boson searches at
the LHC. If the Higgs mass is around 130 GeV, then a promising decay
channel is $H\to ZZ^* \to 4\ell$ where $Z*$ represents an off shell
$Z$. The production of a lepton pair with  one or more heavy quarks
is a background if the heavy quarks decay
leptonically. Fig. \ref{f:dy} shows the relavent diagrams through
order $\alpha_s$; in each case there is a $b$ and $\overline{b}$ in
the event, at order  $\alpha_s^0$ both arise as fragements of the
incoming beams.
\FIGURE{

     \epsfig{file=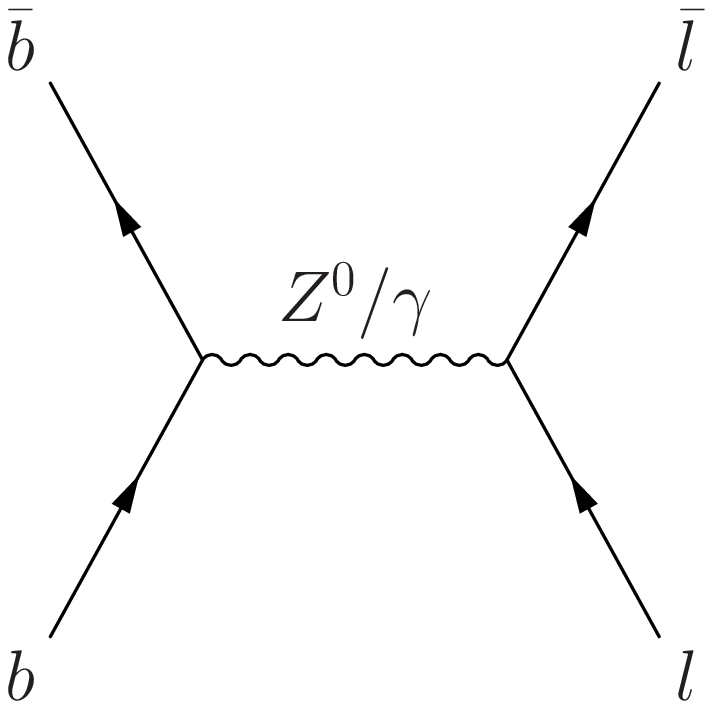,width=4.2cm}\parbox{0.3cm}{\vskip -3.5cm \Large$\oplus$}
     \epsfig{file=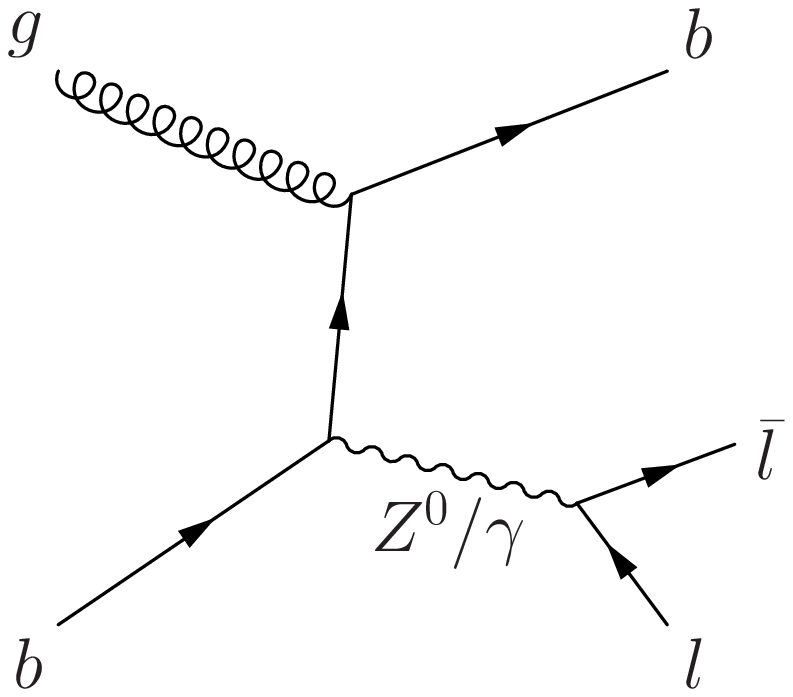,width=4.2cm}\parbox{0.3cm}{\vskip -3.5cm \Large$\ominus$}
     \epsfig{file=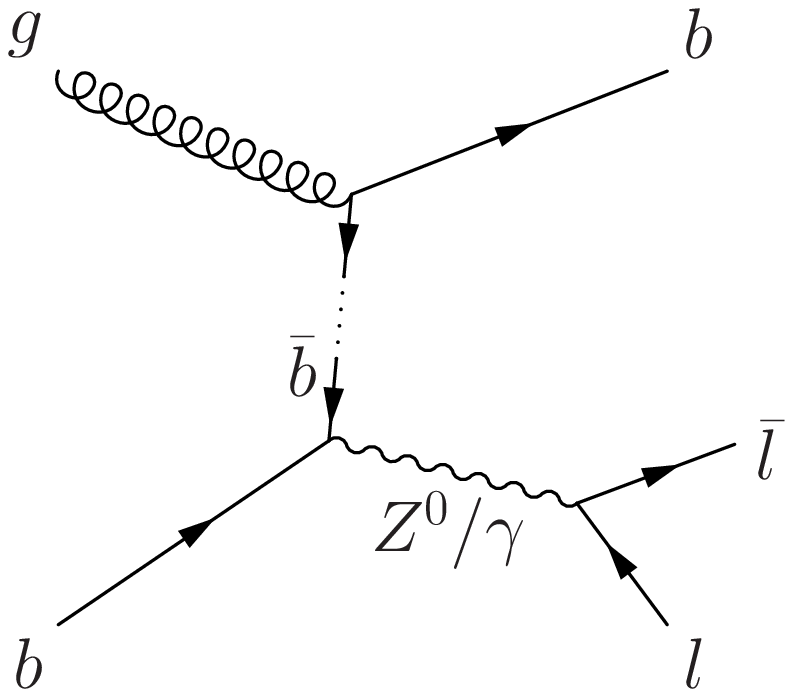,width=4.2cm}

\caption{ Representative Feynman diagrams for the Drell-Yan with
associated b-quark production process for (from left to right): Order $\rm
\alpha_s^{(0)}$, order $\rm \alpha_s^{(1)}$ and order $\rm \alpha_s^{(1)}$ subtraction
term.\label{f:dy}}
}

\TABLE{
\newcommand{\lstrut}{{$\strut\atop\strut$}}
  \caption { The process cross-sections for the leading order
  process $b \bar{b} \to Z \to \mu^+ \mu^-$, next-to-leading order
  process $g b \to Z b \to \mu^+ \mu^- b$ and the subtraction process
  $(g \to b \bar{b}) b \to Z b \to \mu^+ \mu^- b$ in the LHC
  environment (proton-proton collisions at $\rm \sqrt{s}$ = 14 TeV)
  are listed.  The b-quark mass is set to $m_b = 4.8$ GeV and the
  factorization and renormalization scales are set to the $Z^0$
  invariant mass squared. In addition, the order $\rm \alpha_s^{(2)}$
  $gg \to Z b \bar{b} \to \mu^+ \mu^- b \bar{b}$ process cross-section
  is shown for comparison.  The cross-sections are given for the LO
  CTEQ5L \cite{Lai:1999wy} and the derived JCC PDFs. In the Monte-Carlo procedure the
  next-to-leading process weights are combined with the subtraction
  weights on the event-by-event basis as described in the text.
  \label{t:dy}}

\begin{tabular}{lcc}
\hline
Process & $ \sigma_{\mathrm{CTEQ5L,\mu_0 = m_Z}}$ $\rm [ pb ]$ & $ \sigma_{\mathrm{JCC,\mu_0 = m_Z}}$  $\rm [ pb ]$ \\
\hline
$b \bar{b} \to Z \to \mu^+ \mu^-$ & 57.9 & 39.9 \\
\hline
$g b \to Z b \to \mu^+ \mu^- b$ & 72.1  & 60.0 \\
\hline
$(g \to b \bar{b}) b \to Z b \to \mu^+ \mu^- b$ & 73.3  & 60.9 \\
\hline
$\Sigma$ & 56.7 & 39.0 \\
\hline
$gg \to Z  b \bar{b} \to \mu^+ \mu^- b \bar{b}$ & 22.8 & 22.8 \\
\hline
\end{tabular}
 
}

The cross-sections obtained for the leading order process $b \bar{b} \to Z \to
\mu^+ \mu^-$, next-to-leading order process $g b \to Z b \to \mu^+ \mu^-
b$ and the subtraction process $(g \to b \bar{b}) b \to Z b \to \mu^+ \mu^- b$
in the LHC environment (proton-proton collisions at $\rm \sqrt{s}$ = 14 TeV) are
given in the Table \ref{t:dy}, both for leading order PDF (CTEQ5L  \cite{Lai:1999wy} was used) and
the PDF-s evolved according to the Collins prescription (c.f. Equation
\ref{e:jcc}, labeled JCC), along with the cross-section for the order 
$\rm \alpha_s^{(2)}$ $gg \to Z b \bar{b} \to \mu^+ \mu^- b \bar{b}$ process.
Separate cross-section contributions for the next-to-leading order process and
the subtraction term are given for convenience; in the Monte-Carlo procedure
developed in this paper the events are generated according to the differential
cross-section corresponding to the 'hard' order $\alpha_s^{(1)}$ process,
\ie the next-to-leading calculation with the subtraction terms subtracted on an
event-by-event basis.

The differential distributions of the virtuality $\rm \mu$ and the
transverse momentum ($p_T$) distribution of the b-quark with the highest $p_T$ of the two
(produced either in
the hard process or in the subsequent Sudakov showering) are shown in Figure
\ref{f:dydist}, whereby also the $p_T$ distribution of the b-quark with the highest $p_T$
from the order $\rm \alpha_s^{(2)}$ $gg \to Z b \bar{b} \to \mu^+ \mu^- b
\bar{b}$ process is shown.

\FIGURE{

     \epsfig{file=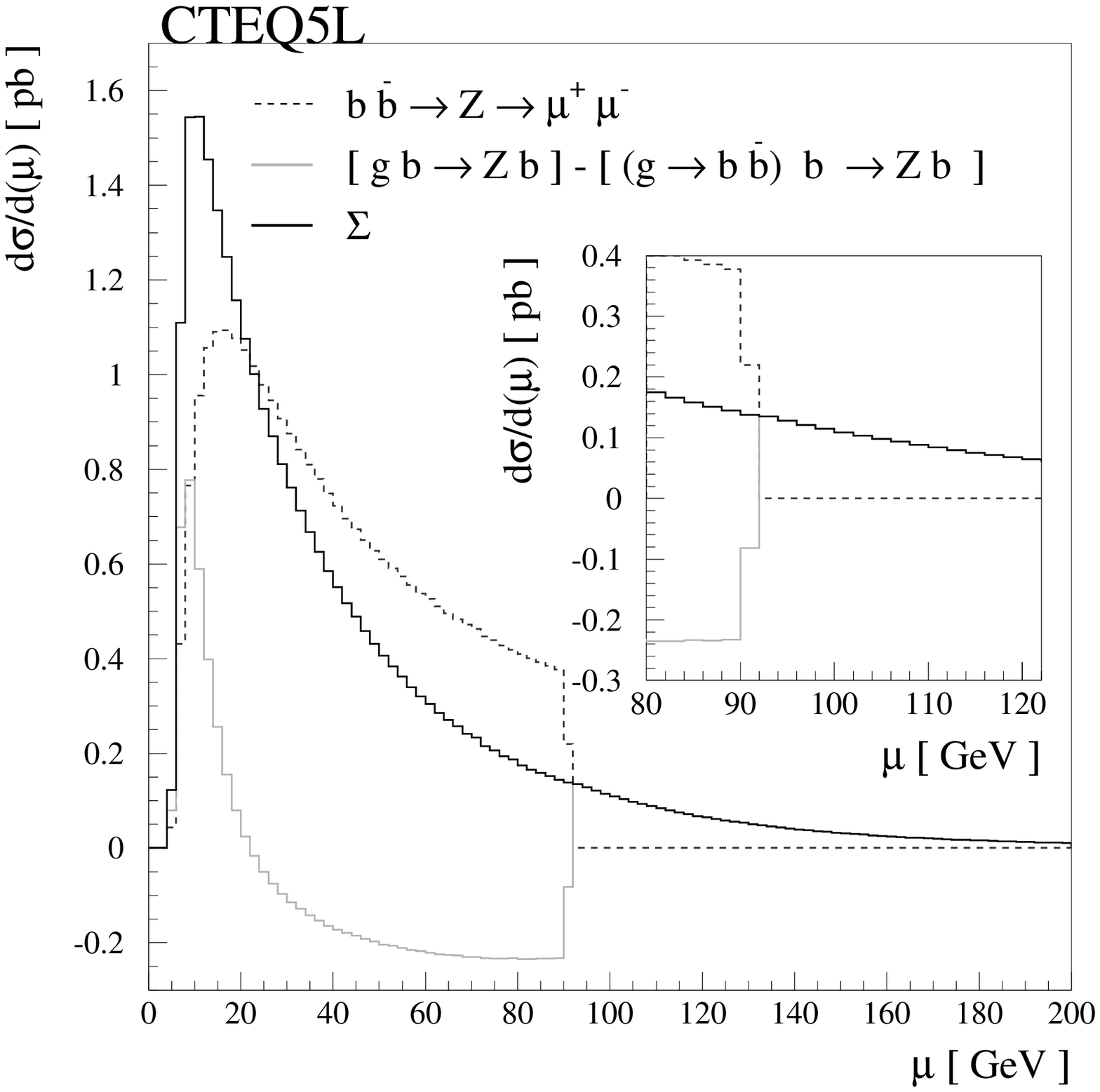,width=0.52\linewidth}\hspace{-1cm}
     \epsfig{file=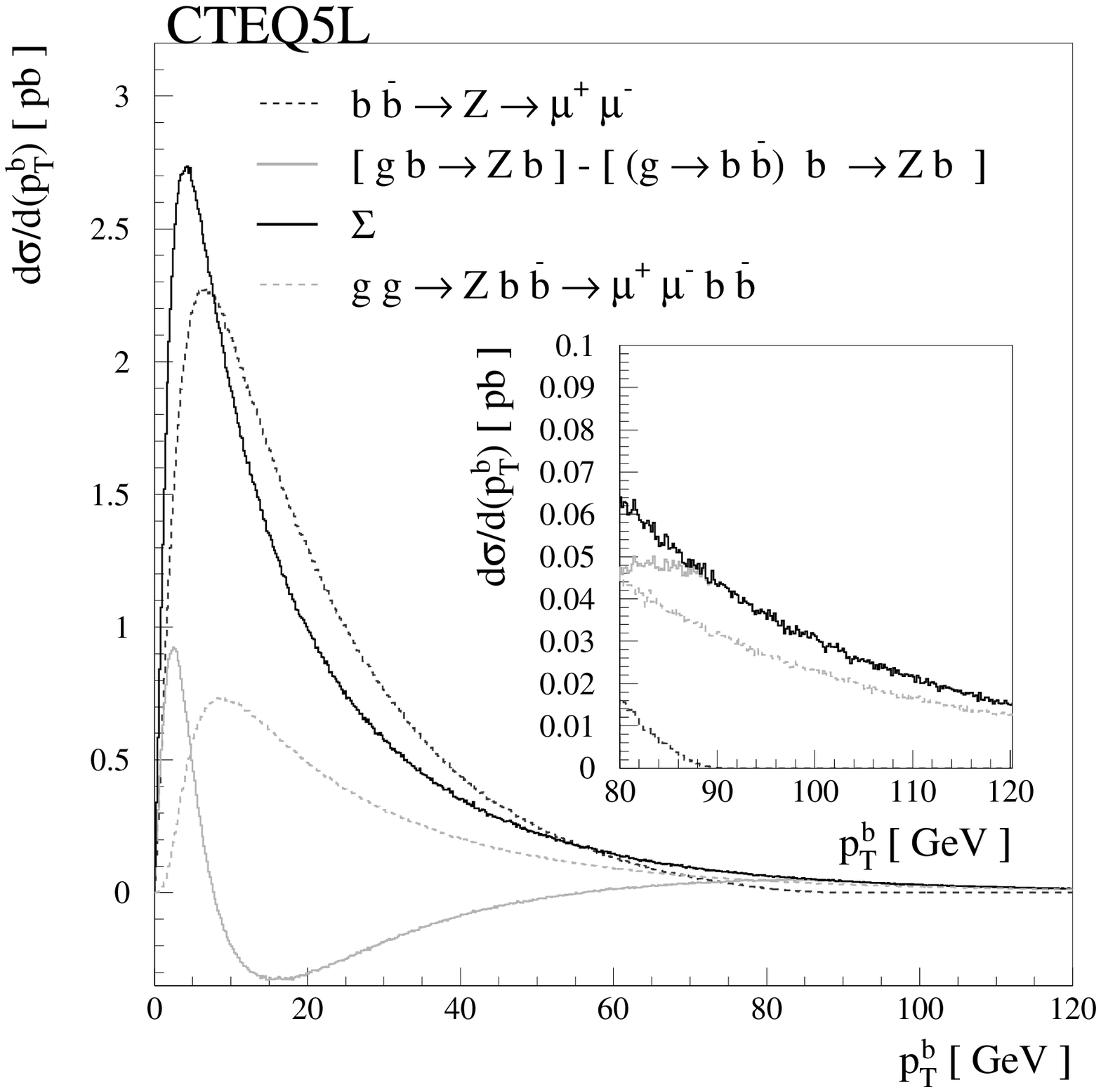,width=0.52\linewidth}

\caption{
 The differential distributions of the virtuality $\rm \mu$ and the transverse
momentum ($p_T$) distribution of the b-quark with the higher virtuality
(produced either in the hard process or in the subsequent Sudakov showering) are
shown for the calculations using the LO CTEQ5L parton density functions. The next-to-leading order process entries contain the subtraction terms
as calculated on event-by event basis.
\label{f:dydist}} 
}

In the results in Fig. \ref{f:dydist} one can  observe a smooth
distribution of the virtuality $\rm \mu$ over the full kinematic range
as the result of the implemented matching procedure. It is 
manifest that the cutoff on the b-quark virtuality $\mu$ and the
resulting subtraction contribution do not map to the $p_T$
distribution in a simple way. An interesting result is that the order
$\rm \alpha_s^{(2)}$ $gg \to Z b \bar{b} \to \mu^+ \mu^- b \bar{b}$
process $p_T$ distribution of the b-quark seems to be quite close to
the result of the merging procedure in the high kinematic range as one
could indeed expect if the perturbative calculations are to be
consistent in the perturbative regime. In the low $p_T$ region the $gg
\to Z b \bar{b} \to \mu^+ \mu^- b \bar{b}$ process undershoots the
expected distribution of the merged $\rm \alpha_s^{(0)} \oplus \rm
\alpha_s^{(1)}$ processes which is to be expected since in this case the
non-perturbative contributions prevail. One has to keep in mind when
comparing the two results results that in the derived $\rm
\alpha_s^{(1)}$ calculation the other incoming b-quark is still
effectively on-shell, \ie its virtuality and branchings are obtained
solely from the Sudakov showering, where as in the $\rm
\alpha_s^{(2)}$ process both incoming b-quarks are treated as
propagators in the full perturbative calculation. A further
improvement would certainly be to repeat and extend the procedures
derived in this paper to include the full order $\rm \alpha_s^{(2)}$
calculation.

From the results one can also see that the use of JCC evolved PDF-s
significantly reduces the cross-section of the leading-order process with
respect to the values obtained using the CTEQ5L \cite{Lai:1999wy} PDFs and to a lesser extent the
cross-sections of the next-to-leading and the subtraction contribution,
since the latter two include only one b-quark and one gluon in the initial state
and  are thus less affected by the change in the b-quark PDF evolution.

\subsection{The 't-channel' Single Top Production Process}

The 't-channel' single top production mechanism is of importance at the LHC since it
provides a clean signal for top quark and W boson polarization
studies. The final state consist of a $t,W$, and $\overline{b}$ as
illustrated in Fig. \ref{f:onet}
\FIGURE{

     \epsfig{file=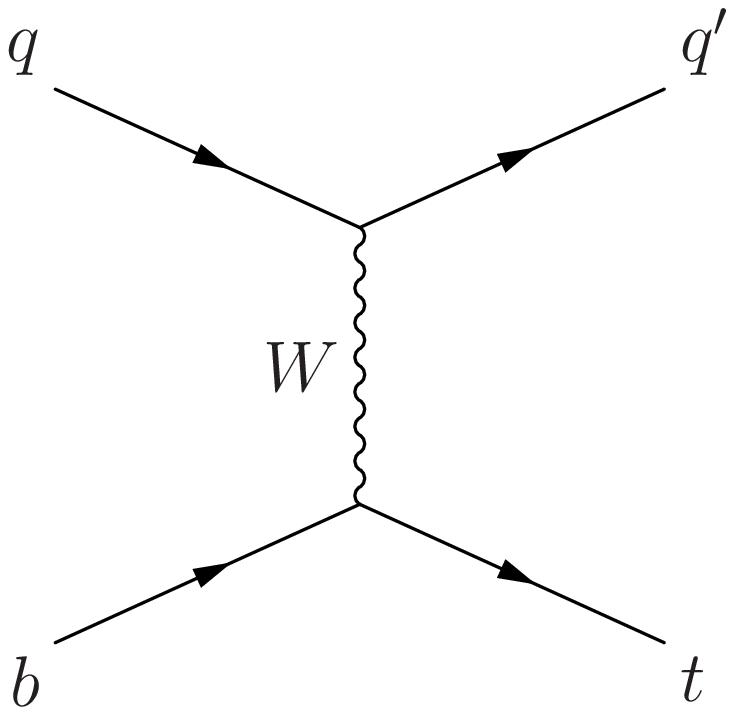,width=4.2cm}\parbox{0.3cm}{\vskip -3.5cm \Large$\oplus$}
     \epsfig{file=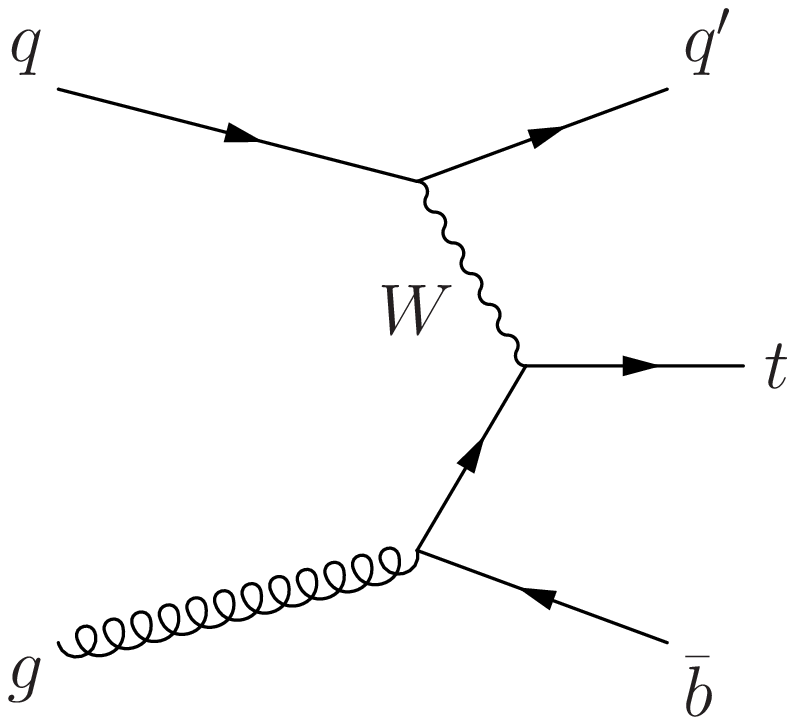,width=4.2cm}\parbox{0.3cm}{\vskip -3.5cm \Large$\ominus$}
     \epsfig{file=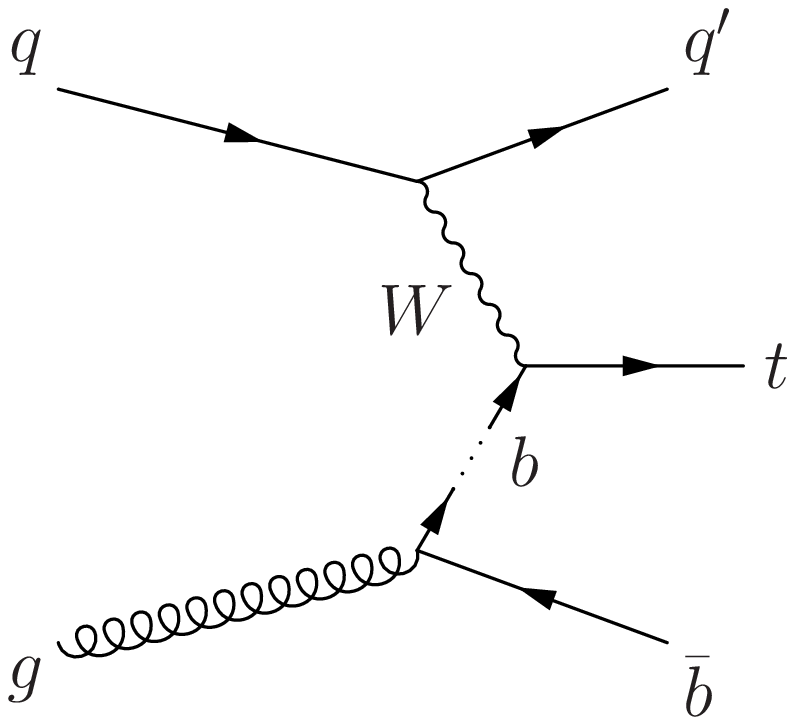,width=4.2cm}

\caption{ Representative Feynman diagrams for the 't-channel' single top production process for (from
left to right): Order $\rm \alpha_s^{(0)}$, order $\rm \alpha_s^{(1)}$ and order $\rm \alpha_s^{(1)}$
subtraction term.\label{f:onet}}
}

The cross-sections obtained for the leading order process $b q \to t q'$,
next-to-leading order process $g q \to t q' \bar{b}$ and the subtraction process
$(g \to b \bar{b}) q \to t q' \bar{b}$ (and charge conjugates) in the LHC
environment (proton-proton collisions at $\rm \sqrt{s}$ = 14 TeV) are given in
the Table \ref{t:onet}, both for leading order PDF (CTEQ5L  \cite{Lai:1999wy} was used) and the
PDF-s evolved according to the Collins prescription (c.f. Equation \ref{e:jcc},
labeled JCC) for the scale choices $\rm \mu_0 = m_t$ and $\rm \mu_0 = 60~GeV$.
The differential distributions of the virtuality $\rm \mu$ and the transverse
momentum ($p_T$) distribution of the b-quark (produced either in the hard
process or in the subsequent Sudakov showering) are shown in Figures
\ref{f:onetdist60} and \ref{f:onetdist}. Separate cross-section contributions are given for 
convenience; in the Monte-Carlo procedure developed in this paper the $g q \to t q' \bar{b}$ events are
generated according to the differential cross-section corresponding to the
'hard' order $\alpha_s^{(1)}$ process, \ie the next-to-leading calculation with
the subtraction terms subtracted on an event-by-event basis.

\TABLE{
\newcommand{\lstrut}{{$\strut\atop\strut$}}
  \caption { The process cross-sections for the leading order
  process $b q \to t q'$, next-to-leading order process $g q \to t q'
  \bar{b}$ and the subtraction process $(g \to b \bar{b}) q \to t q'
  \bar{b}$ (including the charge conjugates) in the LHC environment
  (proton-proton collisions at $\rm \sqrt{s}$ = 14 TeV) are
  listed. These inclusive cross-sections include all top (and $W^\pm$)
  decay channels, the b-quark mass is set to $m_b = 4.8$ GeV and
  top-quark mass to $m_t = 175$ GeV with the factorization and
  renormalization scales set to the top mass values $\rm \mu_0 = m_t$
  and $\rm \mu_0 = 60~GeV$. The cross-sections are given for the LO
  CTEQ5L and JCC PDFs. In the Monte-Carlo procedure the
  next-to-leading process weights are combined with the subtraction
  weights on the event-by-event basis as described in the text.
  \label{t:onet}}

\begin{tabular}{lcccc}
\hline
Process & $ \sigma_{\mathrm{CTEQ5L,\atop \mu_0 = m_t}}$  $\rm [ pb ]$ & $ \sigma_{\mathrm{JCC,\atop \mu_0 = m_t}}$  $\rm [ pb ]$  & $ \sigma_{\mathrm{CTEQ5L,\atop \mu_0 = 60~GeV}}$  $\rm [ pb ]$ & $ \sigma_{\mathrm{JCC,\atop \mu_0 = 60~GeV}}$  $\rm [ pb ]$ \\[8pt]
\hline
$b q \to t q'$ & 222.2 & 187.8 & 178.1 & 138.7  \\
\hline
$g q \to t q' \bar{b}$ & 156.2 & 154.2 & 188.2 & 184.4 \\
\hline
$(g \to b \bar{b}) q \to t q' \bar{b}$ & 140.1 & 138.2 & 102.8 & 100.5 \\
\hline
$\Sigma$ & 238.3 & 203.8 & 263.5 & 222.6 \\
\hline
\end{tabular}
 
}

From the results in Fig. \ref{f:onetdist60} and Fig. \ref{f:onetdist} one can  observe that the
applied procedure produces a very good match of the processes in the combined
distribution of the b-quark virtuality $\mu$ resulting in a smooth
(almost seamless) transition in the vicinity of the cutoff. As one can
also observe the cutoff on the b-quark virtuality $\mu$ and the resulting
subtraction contribution do not map to the $p_T$ distribution in a
trivial manner; hence one can surmise that the simple  methods
involving adding 
of the processes based on $p_T$ distribution cuts  probably give erroneous predictions. 

This procedure can in unmodified form be applied to the full $2 \to 4$ and $2 \to 4$ matrix elements
$b q \to t q' \to W b q' \to f \bar{f}' b q$ and $g q \to t q' \bar{b} \to W b q' \bar{b} \to  f \bar{f}' b q \bar{b}$
including top quark decays and has as such also been implemented in the AcerMC Monte-Carlo generator.
\FIGURE{
     \epsfig{file=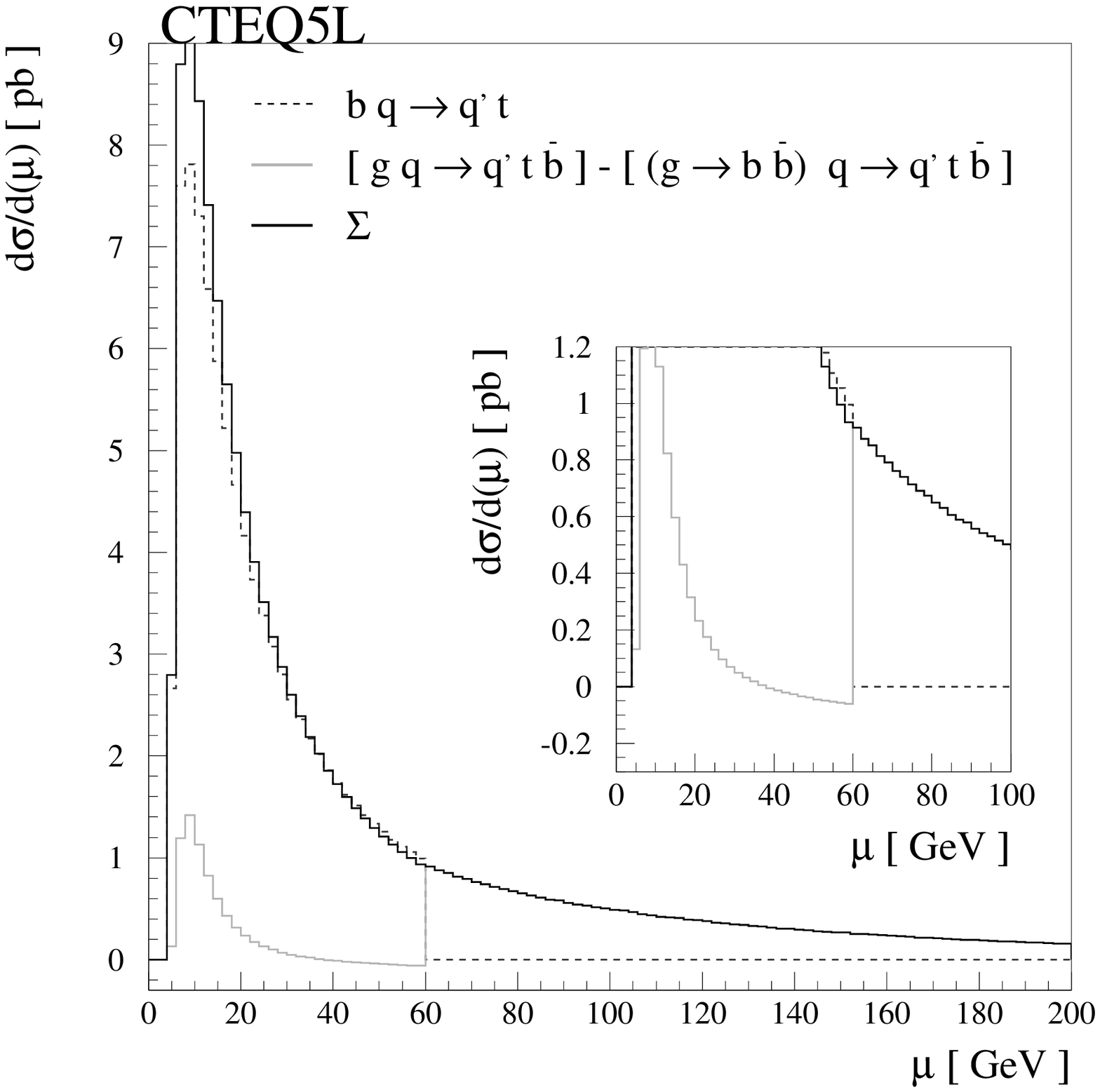,width=0.52\linewidth}\hspace{-1cm}
     \epsfig{file=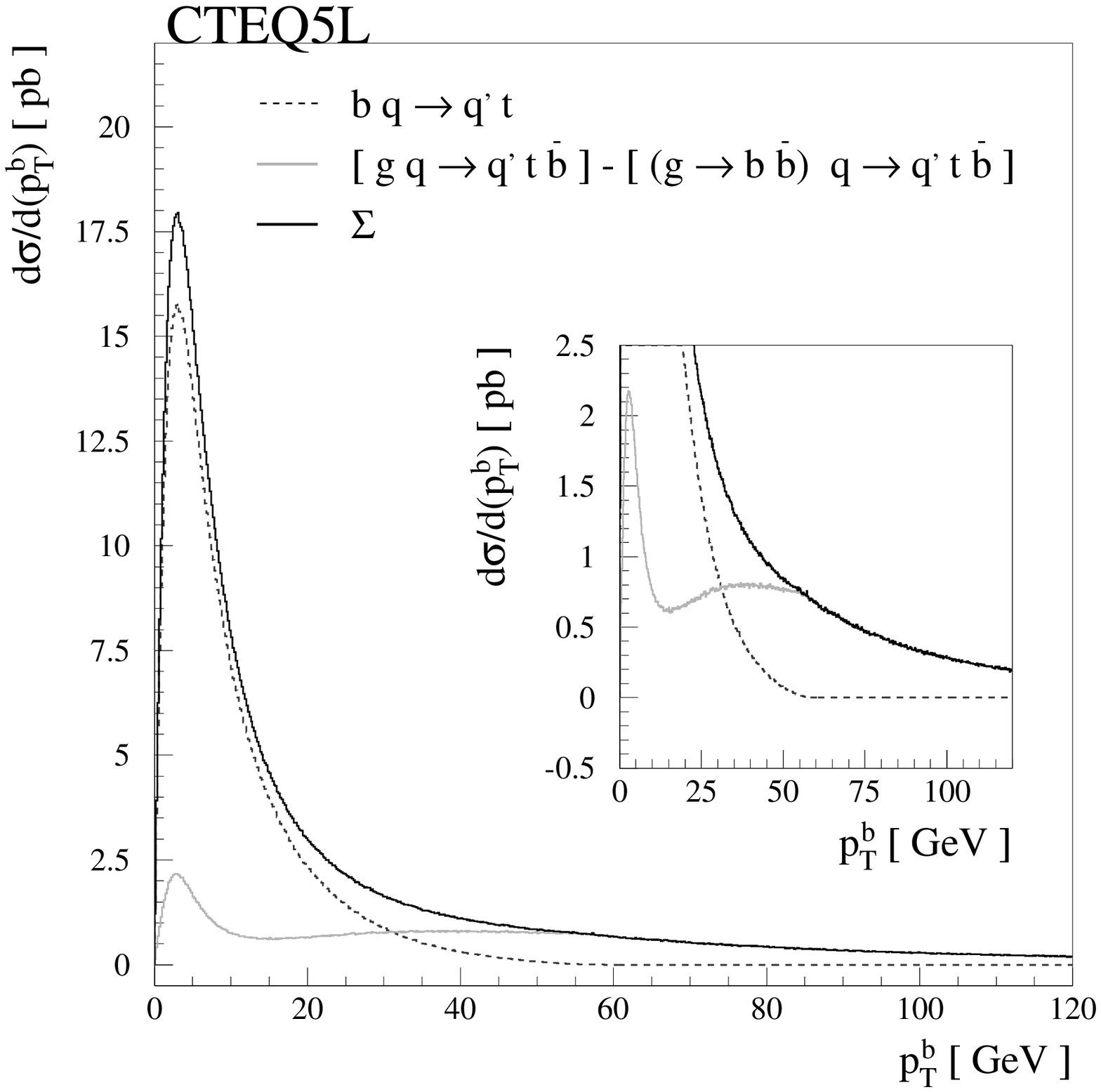,width=0.52\linewidth}
\caption{
 The differential distributions of the virtuality $\rm \mu$ and the transverse
momentum ($p_T$) distribution of the b-quark (produced either in the hard
process or in the subsequent Sudakov showering) are shown for the calculations
using the LO CTEQ5L  PDFs  and the showering
(factorization) scale set to $\rm \mu_0 = 60~GeV$. The next-to-leading order
process entries contain the subtraction terms as calculated on event-by event
basis.
\label{f:onetdist60}} 
}

\FIGURE{
     \epsfig{file=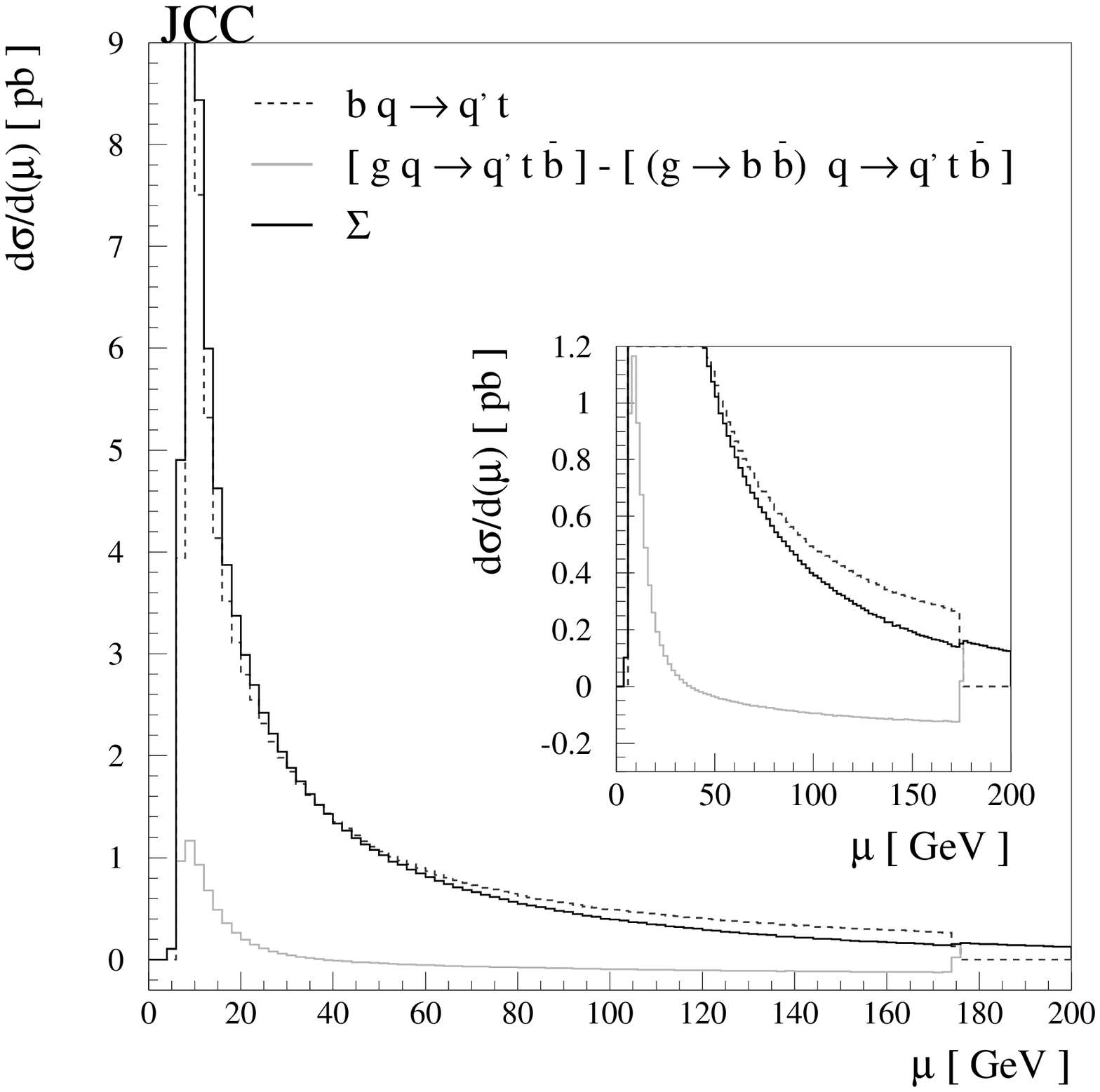,width=0.52\linewidth}\hspace{-1cm}
     \epsfig{file=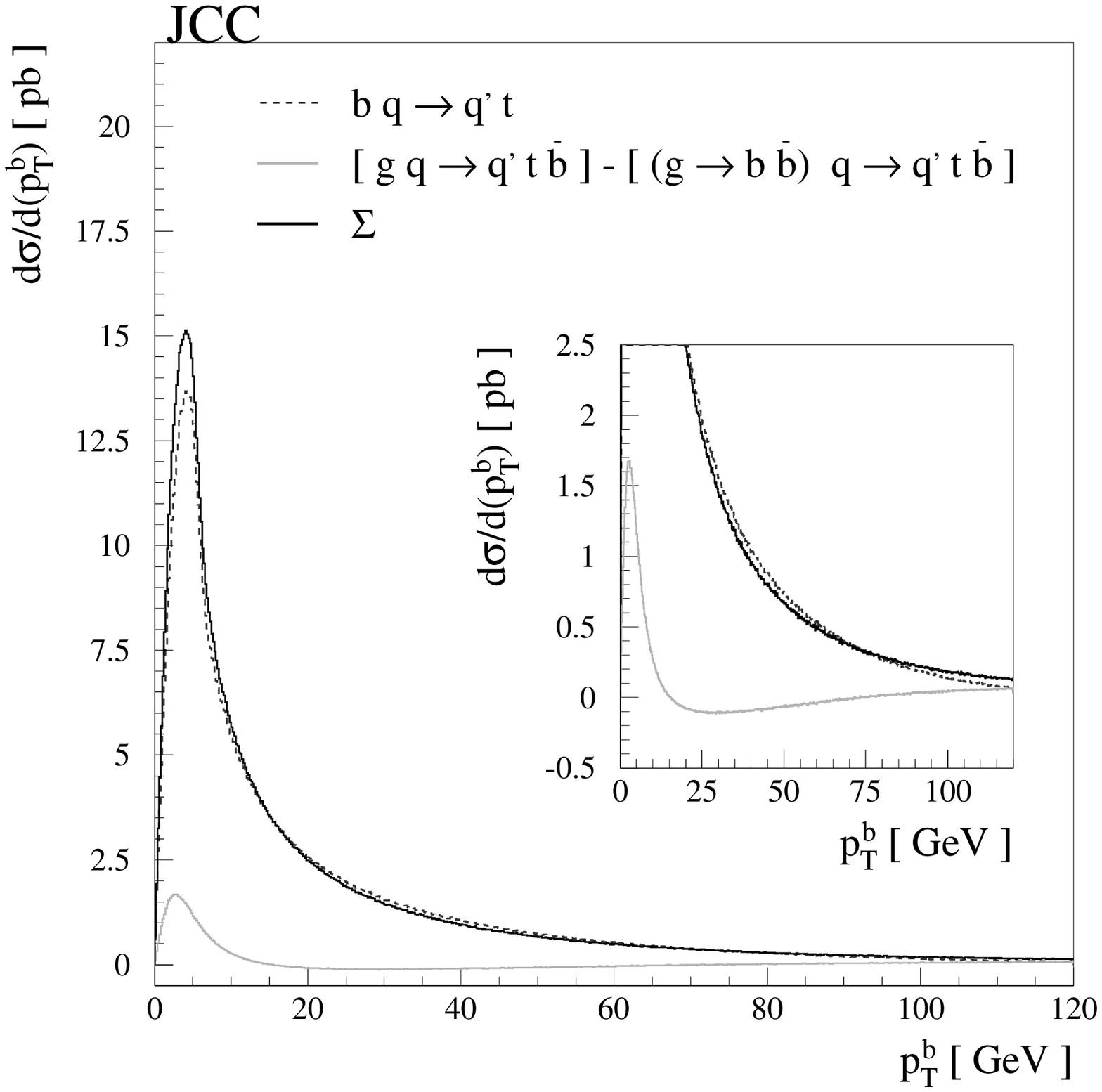,width=0.52\linewidth}
\caption{
 The differential distributions of the virtuality $\rm \mu$ and the transverse
momentum ($p_T$) distribution of the b-quark (produced either in the hard
process or in the subsequent Sudakov showering) are shown for the calculations
using the JCC PDFs  and the showering
(factorization) scale set to $\rm \mu_0 = m_t$. The next-to-leading order
process entries contain the subtraction terms as calculated on event-by event
basis.
\label{f:onetdist}} 
}
\clearpage

One of the interesting results is  that the use of JCC evolved PDF-s
significantly reduces the cross-section of the $b q \to t q'$ process with
respect to the values obtained using the CTEQ5L PDFs and to a lesser extent the
cross-sections of the $g q \to t q' \bar{b}$ and the subtraction contribution,
since the latter two contain only the light quarks and gluons which are less (or
in case of gluons not at all) affected by the change in PDF evolution compared
to the b-quark PDF.

A somewhat cruder method of merging different order processes has for
the 't-channel' single top production already been implemented a while ago in the
program ONETOP\cite{Carlson:1995ei}; the results from the two methods
are compatible within the differences of the methods used in both
implementations. It is instructive to understand  where the differences
between the two procedures originate. Specifically, in ONETOP the
subtraction term is higher than the cross-section of the  $(2 \to 3)$ $g
q \to t q' \bar{b}$ process; from the Fig. 3.6 in
\cite{Carlson:1995ei} the subtraction cross-section of the process $(g
\to b \bar{b}) q \to t q' \bar{b}$ is of the order of about 190 pb compared to the 
140 pb one gets using the procedures described in this paper. The
difference originates in part in the massless approximation of the
participating particles implemented in ONETOP and can be traced back
to the fact that the subtraction term in ONETOP (as given in Appendix
D in \cite{Carlson:1995ei}) is calculated from the \emph{integrated}
parton density function correction (the first-order term in
Eq. \ref{e:collog} of this paper) coupled to the zero-th order $b q \to t q'$
cross-section, since  the virtuality is already integrated over into the
$\log\frac{\mu^2}{m_b^2}$. In contrast, in the present work the procedure
is more complex and requires identifying the virtuality of the $2 \to
3$ process in the massive calculation. In addition, in ONETOP the
spectator energy fraction is kept unchanged whereas in the new
procedure only the rapidity constraint is used instead. The ONETOP
calculation using the integrated PDF correction has been repeated as a
 check and gives a  subtraction term with the
value of about 185 pb which is consistent with the ONETOP results.
 
\subsection{The 'tW channel' Single Top Production Process}

The 'tW-channel' single top production mechanism is also of importance at the LHC since it
provides a clean signal for top quark and W boson polarization
studies. The process is illustrated in Fig. \ref{f:onetw}.
\FIGURE{
     \epsfig{file=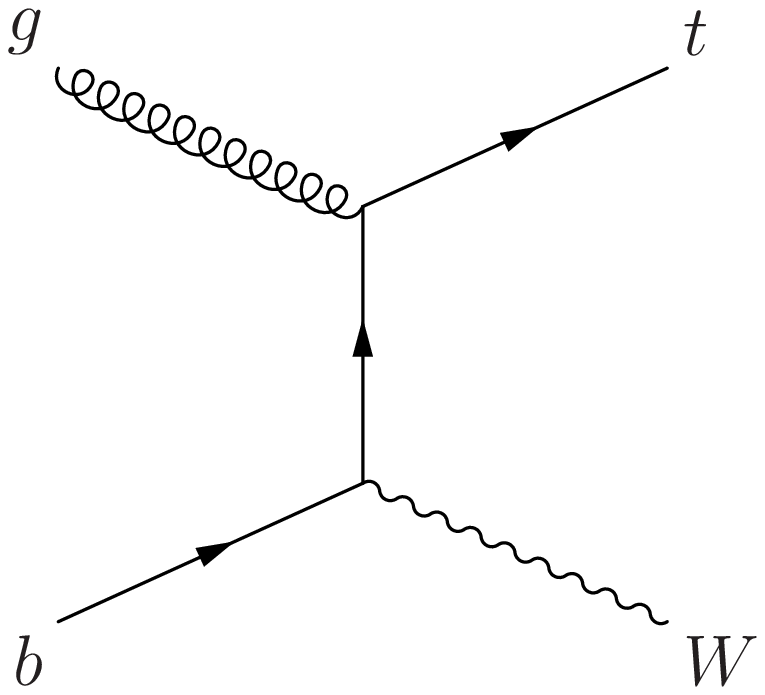,width=4.2cm}\parbox{0.3cm}{\vskip -3.5cm \Large$\oplus$}
     \epsfig{file=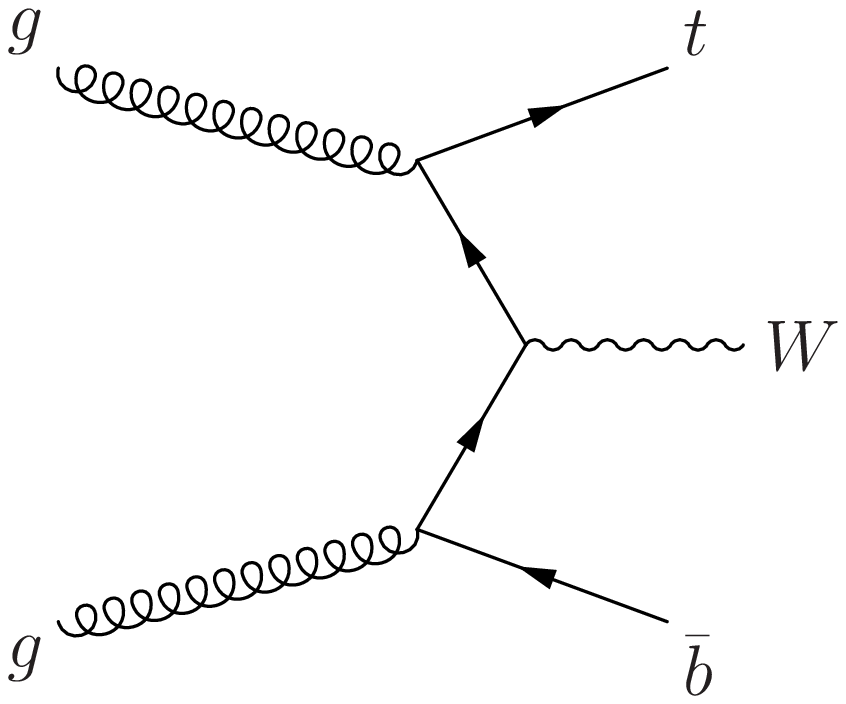,width=4.2cm}\parbox{0.3cm}{\vskip -3.5cm \Large$\ominus$}
     \epsfig{file=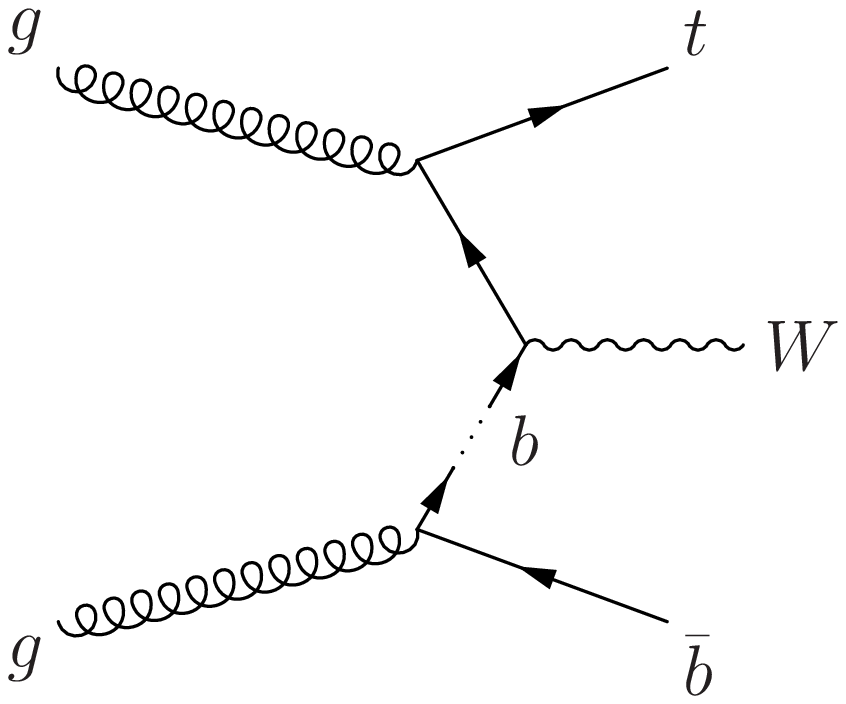,width=4.2cm}
\caption{ Representative Feynman diagrams for the 'tW-channel' single top production process for (from
left to right): Order $\rm \alpha_s^{(0)}$, order $\rm \alpha_s^{(1)}$ and order $\rm \alpha_s^{(1)}$
subtraction term.\label{f:onetw}}
}

The cross-sections obtained for the leading order process $g b \to t (W \to) f \bar{f}'$,
next-to-leading order process $g g \to t  (W \to) f \bar{f}' \bar{b}$ and the subtraction process
$(g \to b \bar{b}) g \to t (W \to) f \bar{f}' \bar{b}$ (and charge conjugates) at the LHC.
are given in
the Table \ref{t:onetw}, both for leading order PDF (CTEQ5L 
 \cite{Lai:1999wy} was used) and the
PDF-s evolved according to the Collins prescription (c.f. Equation \ref{e:jcc},
labeled JCC) for the scale choice $\rm \mu_0 = 60~GeV$.
The differential distributions of the virtuality $\rm \mu$ and the transverse
momentum ($p_T$) distribution of the b-quark (produced either in the hard
process or in the subsequent Sudakov showering) are shown in the Figure
\ref{f:onetwdist60} . Separate cross-section contributions are given for 
convenience; in the Monte-Carlo procedure developed in this paper the
$g g \to t (W \to) f \bar{f}' \bar{b}$ events are generated according
to the differential cross-section corresponding to the 'hard' order
$\alpha_s^{(1)}$ process, \ie\  the next-to-leading calculation with the
subtraction terms subtracted on an event-by-event basis. Note also
that in this case there are indeed two subtraction terms, one
for each incoming gluon.

From the results in Fig. \ref{f:onetwdist60}  one can again observe that the
applied procedure produces a very good match of the processes in the combined
distribution of the b-quark virtuality $\mu$ resulting in a smooth
(almost seamless) transition in the vicinity of the cutoff. As one can
also observe the cutoff on the b-quark virtuality $\mu$ and the resulting
subtraction contribution do again not map to the $p_T$ distribution in a
trivial manner; again therefore one expects  that the simple gluing methods
of the processes based on $p_T$ distribution cuts most probably give erroneous
predictions. 

\TABLE{
\newcommand{\lstrut}{{$\strut\atop\strut$}}
  \caption { The process cross-sections for the leading order
  process 
$g b \to t (W \to) f \bar{f}'$
, next-to-leading order process 
$g g \to t  (W \to) f \bar{f}' \bar{b}$
and the subtraction process 
$(g \to b \bar{b}) g \to t (W \to) f \bar{f}' \bar{b}$
(including the charge conjugates) in the LHC environment
  (proton-proton collisions at $\rm \sqrt{s}$ = 14 TeV) are
  listed. These inclusive cross-sections include all top 
  decay channels, the associated $W^\pm$ decays into muon and neutrino and
   the b-quark mass is set to $m_b = 4.8$ GeV and
  top-quark mass to $m_t = 175$ GeV with the factorization and
  renormalization scales set to the $\rm \mu_0 = 60~GeV$. 
   The cross-sections are given for the LO
  CTEQ5L and JCC PDFs. In the Monte-Carlo procedure the
  next-to-leading process weights are combined with the subtraction
  weights on the event-by-event basis as described in the text.
  \label{t:onetw}}

\begin{tabular}{lcc}
\hline
Process &  $ \sigma_{\mathrm{CTEQ5L,\mu_0 = 60~GeV}}$  $\rm [ pb ]$ & $ \sigma_{\mathrm{JCC,\mu_0 = 60~GeV}}$  $\rm [ pb ]$ \\
\hline
$g b \to t (W \to) \mu \bar{\nu}_\mu $ & 5.9  & 4.7  \\
\hline
$g g \to t  (W \to) \mu \bar{\nu}_\mu  \bar{b}$& 6.0 & 6.0    \\
\hline
$(g \to b \bar{b}) g \to t (W \to) \mu \bar{\nu}_\mu  \bar{b}$ & 3.1  & 3.1 \\
\hline
$\Sigma$ &  8.8  &  7.6 \\
\hline
\end{tabular}
 
}

The diagrams of the process  $gg \to tWb \to W W b \bar b \to f f f f b \bar b$ are in fact just
a subset of 31 Feynman diagrams which have a $W W b \bar b$ intermediate state
(and which also include the $t \bar{t}$ production). Accordingly, the derived
subtraction procedure has in, AcerMC, been applied to the processes having the
full set of Feynman diagrams and the above plots and values should be
considered only  as the validation of the procedure in case of the 'tW channel' single
top production.   
\clearpage 

\FIGURE{
     \epsfig{file=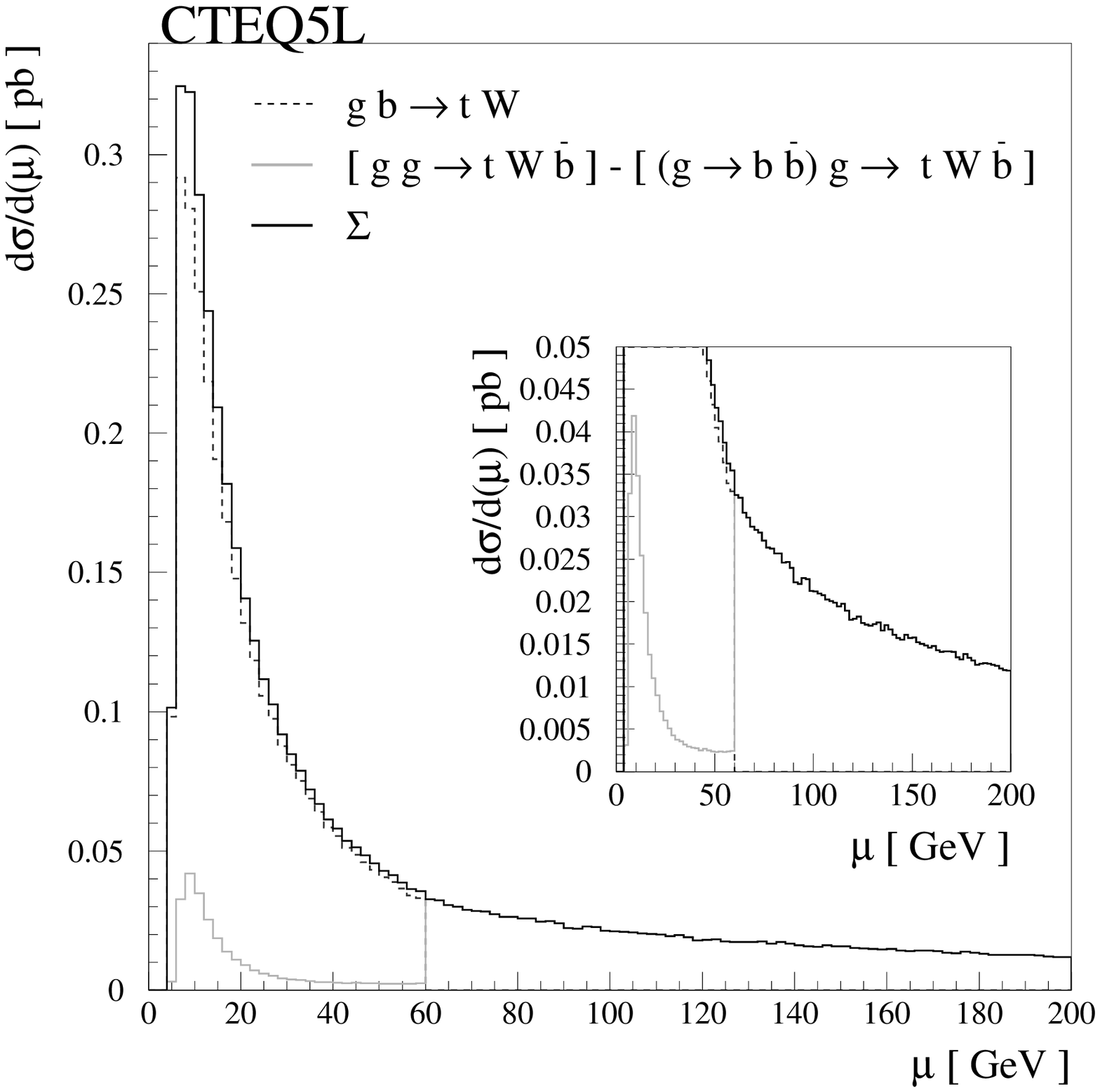,width=0.52\linewidth}\hspace{-1cm}
     \epsfig{file=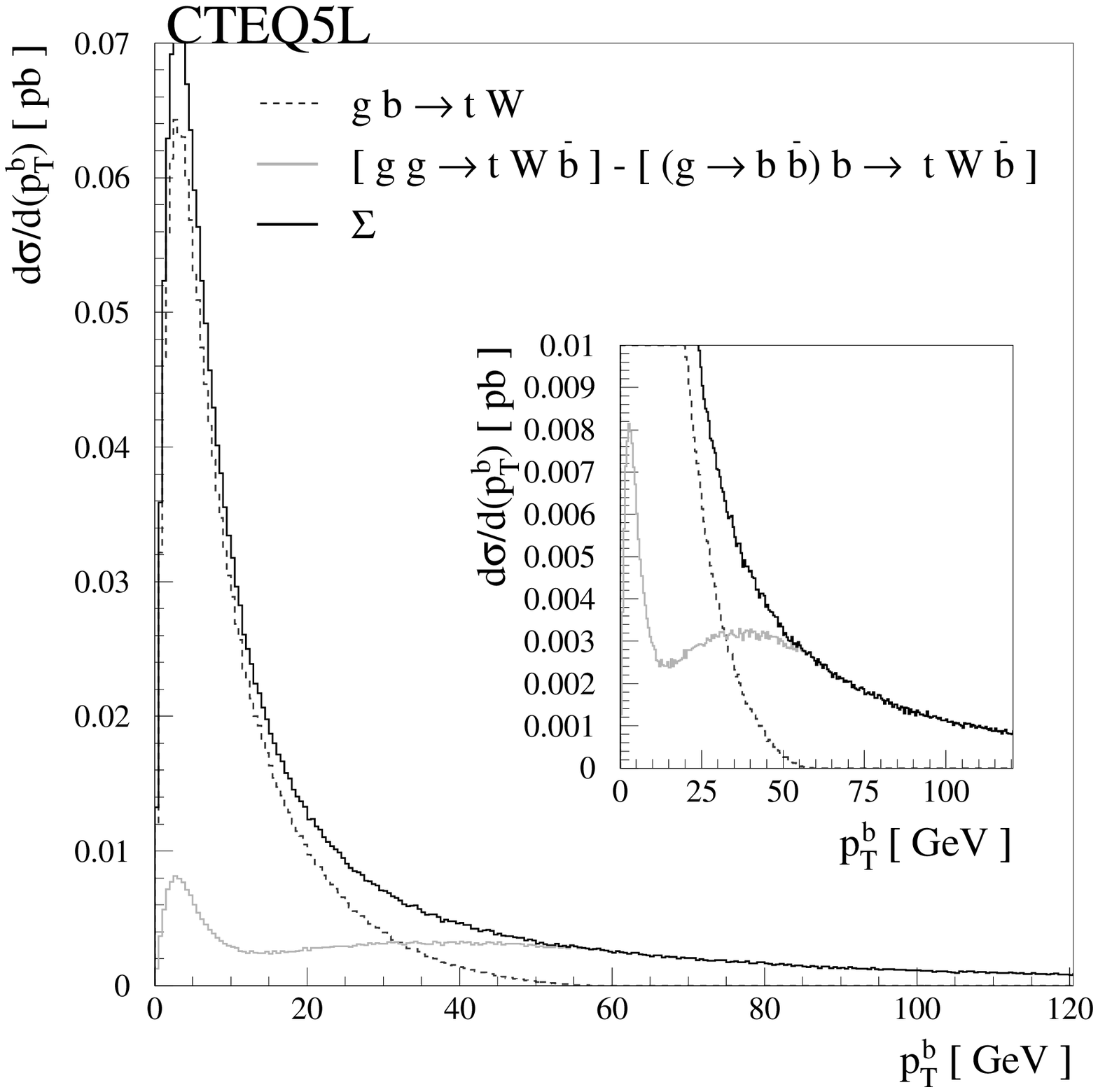,width=0.52\linewidth}
\caption{
 The differential distributions of the virtuality $\rm \mu$ and the transverse
momentum ($p_T$) distribution of the b-quark (produced either in the hard
process or in the subsequent Sudakov showering) are shown for the calculations
using the LO CTEQ5L  PDFs  and the showering
(factorization) scale set to $\rm \mu_0 = 60~GeV$. The next-to-leading order
process entries contain the subtraction terms as calculated on event-by event
basis.
\label{f:onetwdist60}} 
}

\section{Conclusion}

It has been demonstrated explicitly how to deal with the case where a particle of interest can be produced 
from a partonic hard scattering process or as a remnant of an incoming
hadron beam. Examples of the proceedure have been provided and
contrasted with the more {\it ad-hoc} proceedures used previously to
prevent double counting.

\acknowledgments
BPK would like to than Elzbieta Richter-Was for many fruitful discussions on the topic of this paper.
The work of IH was supported  by the Director, Office of Science, Office of
High Energy Physics, of the U.S.\ Department of Energy under Contract
DE-AC02-05CH11231.

\appendix
\section{Kinematic Relations\label{app:conditions}}

The requirement ({\bf 2}) in the list of Subsection \ref{s:npart} gives the equivalence
$\rm  \xi_a \equiv x_a$. The remaining relations between  $\rm \xi_c,\xi_b$ and $x_b,s_{n-1}$ are then given
by energy and rapidity conservation requirements and are thus given by the
conditions:
\begin{equation}
s_{n-1} = (p_c + p_b)^2 = m_c^2 + m_b^2 + 2 (p_c^+ p_b^- + p_c^- p_b^+) = 
m_c^2 + m_b^2 + \xi_c \xi_b s + \frac{m_c^2 m_b^2}{\xi_c \xi_b s}
\end{equation}
where $p_c=(p_c^+,\vec{0}^T,p_c^-)=(\xi_c P_A^+,\vec{0}^T,\frac{m_c^2}{2 \xi_c P_A^+})$ 
and $p_b = (p_b^+,\vec{0}^T,p_b^-)=(\frac{m_b^2}{2 \xi_b P_B^-},\vec{0}^T,\xi_b
P_B^-)$ and:
\begin{equation}
y = \frac{1}{2} \ln\left( \frac{k_{n-1}^+}{k_{n-1}^-} \right) =  
\frac{1}{2} \ln\left( \frac{\xi_c}{\xi_b} \right) 
+ \frac{1}{2} \ln\left( \frac{ \xi_c \xi_b s + m_b^2}{ \xi_c \xi_b s + m_c^2 } \right)
\label{e:ynm1}
\end{equation}
with:
\begin{equation}
\frac{k_{n-1}^+}{k_{n-1}^-} = \left(\frac{1+\beta}{1-\beta}\right)
\left(\frac{k_{n-1}^{+*}}{k_{n-1}^{-*}}\right)~~~~~~~\left(\frac{1+\beta}{1-\beta}\right)
=\frac{x_a (x_a x_b s + m_b^2)}{x_b (x_a x_b s + m_a^2)}
\end{equation}
and:
\begin{equation}
k_{n-1}^{\pm\star} = \frac{1}{\sqrt{2}} \left[ \frac{s_n + s_{n-1} - m_{\bar{c}}^2}{2\sqrt{s_n}} \mp \frac{t_{n-1}
- m_b^2 - s_{n-1} + 2 \left(\frac{s_n + m_b^2 - m_a^2}{2\sqrt{s_n}} \right)\left(  \frac{s_n + s_{n-1} -
m_{\bar{c}}^2}{2\sqrt{s_n}} \right)}{2 \left(\frac{\sqrt{\lambda(s_n,m_a^2,m_b^2)}}{2\sqrt{s_n}}\right)} \right]
\end{equation}
which can be inverted to give the expressions for $\xi_c$ and $\xi_b$ as
functions of $s_n,x_b,\ldots$ A further simplification in derivation can be
achieved by introducing another set of variables $\bar{\tau} = \xi_c \cdot
\xi_b$ and $\bar{y} = 1/2 \ln (\xi_c/\xi_b)$ with the Jacobian of the transformation
$\mathcal{J}\frac{(\xi_c,\xi_b)}{(\bar{\tau},\bar{y})} = 1$ and subsequently:
\begin{eqnarray}
\bar{\tau}& = &\frac{1}{2 s} \left\{ \left( s_{n-1} -(m_c^2 + m_b^2) \right) + 
\sqrt{\lambda(s_{n-1},m_c^2,m_b^2)} \right\}, \\
\bar{y} & = & \frac{1}{2} \ln\left( \frac{k_{n-1}^+}{k_{n-1}^-} \right) - 
\frac{1}{2} \ln\left( \frac{ \bar{\tau} s + m_b^2}{ \bar{\tau} s + m_c^2 } \right).
\end{eqnarray}
As one can observe the $\bar{\tau}$ is only a function of $s_{n-1}$ so the only
remaining term to compute in the Jacobian of the transformation
$\mathcal{J}\frac{(\bar{\tau},\bar{y})}{(s_{n-1},x_b)}$ is $d\bar{y}/dx_b$:
\begin{equation}
\mathcal{J}\frac{(\bar{\tau},\bar{y})}{(s_{n-1},x_b)}  = 
\left|\frac{d\bar{\tau}}{d s_{n-1}} \right| \cdot 
\left| \frac{d\bar{y}}{d x_{b}} \right| = 
 \mathcal{F}(s_n,s_{n-1},t_{n-1},x_a,x_b,m_a,m_b,m_{\bar{c}})
\end{equation}
with $\mathcal{F}$ being a lengthy function of the listed parameters and
therefore omitted. In the massless approximation the above expression reduces
to:
\begin{equation}
\mathcal{J}\frac{(\xi_c,\xi_b)}{(s_{n-1},x_b)}_{m\to 0} = 
\left| \frac{x_a}{2 (x_a x_b s + t_{n-1})} + \frac{1}{2 x_b s} \right|
\end{equation}
which is in agreement with the expressions derived by Chen, Collins \emph{et al.} 
\cite{Collins:2000qd,Chen:2001nf,Chen:2001ci,Collins:2002ey}. Analogous
expressions can trivially be obtained also for the split of the other parton.

\end{document}